\documentclass[a4paper,fleqn]{cas-sc}

\usepackage[authoryear]{natbib}
\usepackage{graphicx}
\usepackage{dcolumn}
\usepackage{bm}
\usepackage{siunitx}
\usepackage{amssymb}
\usepackage{subcaption}
\usepackage{soul}
\usepackage{hyperref}
\usepackage{color,soul}

\def\tsc#1{\csdef{#1}{\textsc{\lowercase{#1}}\xspace}}
\tsc{WGM}
\tsc{QE}

\begin{document}
\let\WriteBookmarks\relax
\def\floatpagepagefraction{1}
\def\textpagefraction{.001}

\shorttitle{Dislocation density transients and saturation in irradiated Zr}    

\shortauthors{A. R. Warwick, R. Thomas \textit{et al}.}  

\title[mode = title]{Dislocation density transients and saturation in irradiated zirconium}  



%
\author[inst1]{Andrew R. Warwick}
\cormark[1]
\ead{andrew.warwick@ukaea.uk}

\author[inst2]{Rhys Thomas}
\author[inst1]{M. Boleininger}

\author[inst2]{Ö. Koç}
\author[inst1]{G. Zilahi}
\author[inst5]{G. Ribárik}
\author[inst6]{Z. Hegedues}
\author[inst6]{U. Lienert}
\author[inst2,inst5]{T. Ungar}
\author[inst2]{C. Race}
\author[inst2,inst7]{M. Preuss}
\author[inst2]{P. Frankel}

\author[inst1]{S. L. Dudarev}

\affiliation[inst1]{organization={UK Atomic Energy Authority},
            addressline={Culham Science Centre}, 
            city={Abingdon},
            postcode={OX14 3DB}, 
            country={UK}}

\affiliation[inst2]{organization={Department of Materials, University of Manchester},
            city={Manchester},
            postcode={M13 9PL}, 
            country={UK}}


\affiliation[inst5]{organization={Department of Materials Physics, E\"otv\"os University},
            postcode={PO Box 32, H-1518},
            city={Budapest},
            country={Hungary}}          

\affiliation[inst6]{organization={Deutsches Elektronen-Synchrotron DESY},
            addressline = {Notkestr. 85},
            postcode={22607},
            city={Hamburg},
            country={Germany}}    
            
\affiliation[inst7]{organization={Monash University}, 
city={Clayton}, 
postcode ={VIC 3800}, 
country = {Australia}}

\cortext[1]{Corresponding author}



\begin{abstract}
Zirconium alloys are widely used as the fuel cladding material in pressurised water reactors, accumulating a significant population of defects and dislocations from exposure to neutrons. We present and interpret synchrotron microbeam X-ray diffraction measurements of proton-irradiated Zircaloy-4, where we identify a transient peak and the subsequent saturation of dislocation density as a function of exposure. This is explained by direct atomistic simulations showing that the observed variation of dislocation density as a function of dose is a natural result of the evolution of the dense defect and dislocation microstructure driven by the concurrent generation of defects and their subsequent stress-driven relaxation. In the dynamic equilibrium state of the material developing in the high dose limit, the defect content distribution of the population of dislocation loops, coexisting with the dislocation network, follows a power law with exponent $\alpha \approx 2.2$. This corresponds to the power law exponent of $\beta \approx 3.4$ for the distribution of loops as a function of their diameter that compares favourably with the experimentally measured values of $\beta $ in the range $3\le \beta \le 4$.
\end{abstract}


\begin{highlights}
\item Dislocation density evolution in irradiated Zr has been predicted and measured.
\item A transient peak and subsequent saturation in dislocation density has been observed.
\item Dislocation loop diameters in heavily irradiated Zr are power law distributed.
\end{highlights}

\begin{keywords}
 zirconium \sep irradiation \sep dislocations \sep defects
\end{keywords}

\maketitle
\section{Introduction}

In the core of modern boiling (BWR) or pressurized (PWR) water reactors, the uranium dioxide fuel assemblies are immersed in circulating pressurised water and thus it is critical that only the heat produced by the fission reactions is transported by the coolant and there is no contamination of the coolant from the radioactive fuel itself. %
Hence, the fuel is cladded to protect the reactor environment from contamination, be that during reactor operation or in transit. %
From the design choices that date back over fifty years \citep{Rickover1975}, zirconium alloys are currently employed as the uranium dioxide fuel cladding in water-cooled reactors. %
Containing more than 95 wt\% Zr, these alloys are mostly pure zirconium, chosen for its low neutron absorption cross section \citep{Pomerance1951}. %
Small amounts of Sn, Nb, Fe and/or Cr in the alloys help protect against corrosion and improve structural integrity \citep{Lemaignan2012,Onimus2012}. %

\begin{figure}
    \centering
    \includegraphics[width = 0.8
    \textwidth]{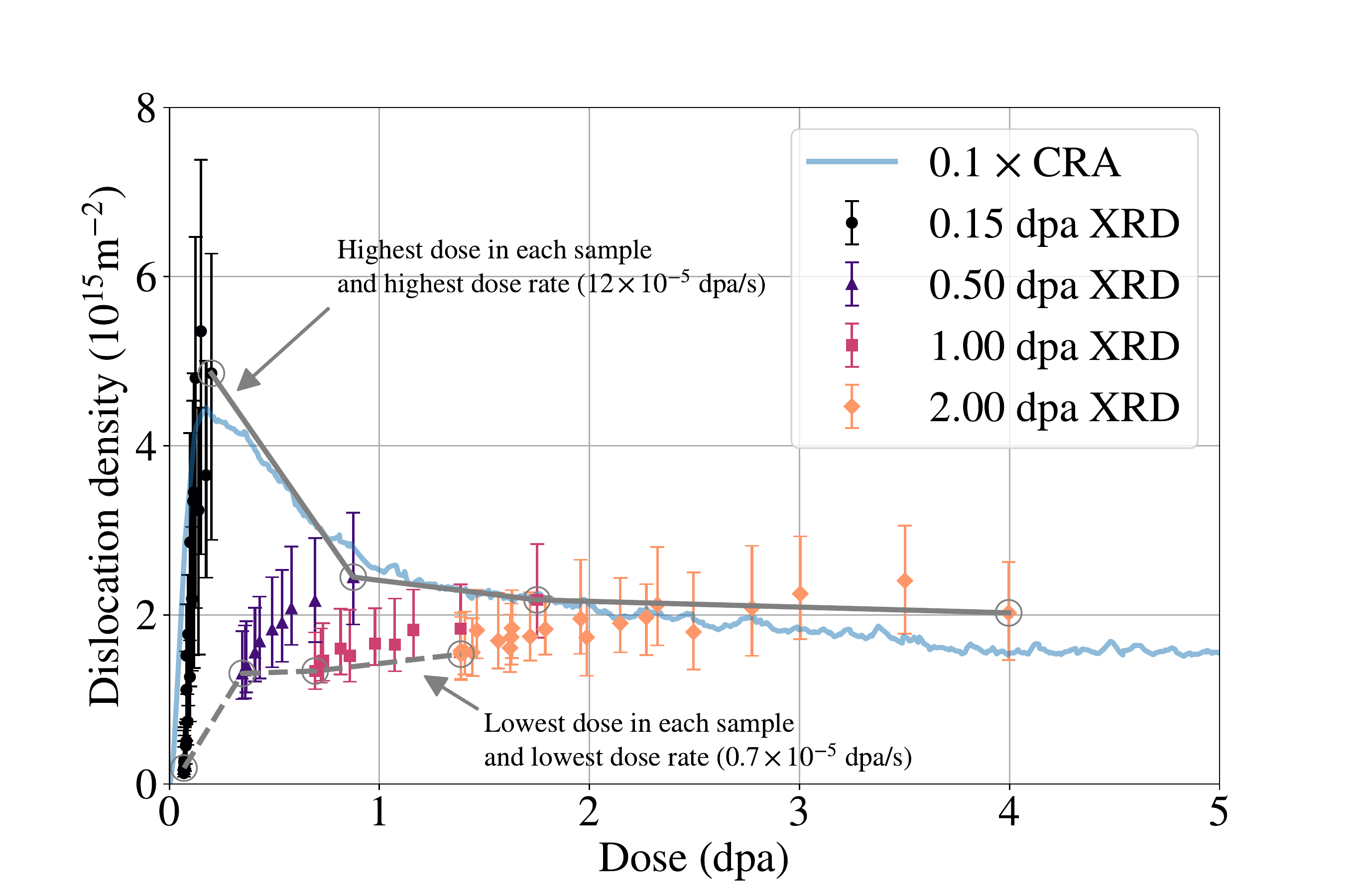}
    \caption{Dislocation density as a function of dose observed experimentally using microbeam synchrotron X-ray diffraction (XRD) measurements of proton-irradiated Zircaloy-4, compared to predictions derived from simulations of pure $\alpha$-Zr performed using the creation relaxation algorithm technique. Four experimental samples with nominal doses of 0.1, 0.5, 1.0 and 2.0 dpa were scanned across the variable dose regime. The simulation results are scaled by a factor of 0.1 and averaged over three different interatomic potentials (see text).}
    \label{fig:CRA_XRD_comparison}
\end{figure}

The elastic neutron scattering cross-section for a Zr nucleus is, however, similar to that of other elements in the Periodic table \citep{Sears2006}, and a prolonged exposure to neutron irradiation results in the accumulation of a considerable amount of microscopic radiation defects generated from the atomic recoils initiated by collisions with neutrons. %
This gives rise to the deterioration of mechanical and physical properties and stimulates dimensional changes \citep{Holt1988,Onimus2012}. %
The high energy \SI{>1}{MeV} neutrons produced by the fissile uranium oxide fuel \citep{Nicodemus1953} collide with Zr nuclei, initiating collision cascades that displace atoms from their lattice sites, rearrange the crystal structure, and generate crystal lattice defects \citep{Domain2005}. %
Defects accumulate with increasing exposure to neutron irradiation, in the form of pairs of self-interstitial and vacancy defects (Frenkel pairs) as well as in the form of clusters of defects that eventually coalesce into large-scale defects, such as dislocation loops and dislocations \citep{Warwick2021}. %
At temperatures above 300-\SI{350}{\degreeCelsius} where point defect diffusion occurs at an appreciable rate comparable or faster rates than the dose rate, it is important to consider random Brownian motion of defects to interfaces, grain boundaries, dislocations or other point defect clusters, whereas at lower temperatures the microstructure of accumulating defects is dominated by other factors.
Given the significance of the life-limiting effect of structural changes on the properties of zirconium cladding, there is significant research effort aimed at improving the performance of structural zirconium alloys in the operating environment of a fission reactor \citep{Adamson2019,Zinkle2013}. %

The exposure of a material to energetic particles is often quantified by the notion of {\it dose}, typically expressed in the units of `displacement per atom' (dpa). Dpa is a simple measure of exposure of a material to radiation and represents the average number of times each atom in the material has been a part of a Frenkel self-interstitial-vacancy defect pair. %
Typically, a zirconium alloy cladding is exposed to \SI{\sim 15}{dpa} over the five years of service \citep{Zinkle2013}. %
Producing an accurate safety case involves the identification of the microstructure formed at a given dose, temperature, and externally applied load. %

In particular, the formation of dislocation loops and dislocations is known to play a critical role in the resulting deterioration of the cladding's structural and mechanical properties. %
For instance, in the large dose limit, high densities of defects and dislocations accumulate in the cladding, causing embrittlement. %
Another important degradation mode is the so-called `irradiation-induced growth' (IIG) that arises from the anisotropy of the hexagonal close packed (hcp) crystal structure of zirconium ($\alpha$-Zr) and zirconium alloys \citep{Griffiths2020, Onimus2022}. %
This crystal structure is stable up to and beyond the reactor core's operating temperature range of \SI{280}{\degreeCelsius} to \SI{350}{\degreeCelsius}, exhibiting an hcp-bcc instability only at significantly higher temperatures above \SI{860}{\degreeCelsius} \citep{Willaime1989}. %

With increasing dose, the interstitial and vacancy-type dislocation loops with the $ \frac{1}{3} \langle 2 \bar{1} \bar{1} 0 \rangle $ Burgers vectors, inhabiting the prismatic crystallographic planes, form a population of the so-called `\textit{a} loops'. %
At relatively low temperatures, this strongly correlates with observed elongation along the `\textit{a}' $\langle 2 \bar{1} \bar{1} 0 \rangle$ and contraction along the `\textit{c}' $\langle 0 0 0 1 \rangle$ crystallographic directions, saturating  at doses larger than \SI{1}{dpa} \citep{Holt1988}. %
At temperatures above \SI{\sim 300}{\degreeCelsius} and doses exceeding \SI{\sim 5}{dpa}, large vacancy-type `\textit{c} loops' with Burgers vectors $ \frac{1}{2} \langle 0 0 0 1 \rangle $ or $ \frac{1}{6} \langle 2 0 \bar{2} 3 \rangle $ appear in the basal crystallographic planes. %
Accompanying this onset of formation of the \textit{c}-type vacancy loops, the magnitudes of \textit{a} and \textit{c} strains increase linearly with dose in a phenomenon called `the breakaway growth' \citep{CHOI2013}. %
Whilst it is known that dislocations have substantial elastic relaxation volume \citep{Dudarev2018,Boleininger2022} giving rise to significant dimensional changes \citep{Christensen2020}, the present understanding of this IIG is mostly phenomenological and existing models are unable to predict, from first principles, the variation of the dislocation content consistent with, or at least verified by, the experimental data. %

Below we present experimental observations and predictions derived from simulations, showing that the density of dislocations saturates in the temperature and dose range where dimensional changes exhibit saturation. %
This is summarised in Figure \ref{fig:CRA_XRD_comparison} illustrating the dislocation densities experimentally measured in proton-irradiated Zircaloy-4 (Zr-1.5\%Sn-0.2\%Fe-0.1\%Cr) together with the simulation data for irradiated $\alpha$-Zr plotted as a function of dose. %
In agreement with observations performed using ion irradiation \citep{Yu2017}, we find that the density of dislocations evolves through a transient at doses \SI{<1}{dpa} before saturating at larger doses. %
The microstructures produced by our simulations indicate that at very low doses \SI{<<1 }{dpa}, small dislocation loops form, which subsequently grow and coalesce into a complex interconnected dislocation network developing at moderate doses \SI{<1}{dpa}. The dislocation network eventually forms complete crystallographic planes \citep{Mason2020,Boleininger2023} and partly dissociates into large dislocation loops at high doses \SI{\gg 1}{dpa}. %
The formation of dislocation loops and dislocations is principally driven by the stress-mediated clustering of self-interstitial atoms, where in the high dose limit the self-interstitial cluster population size distribution follows a power law $p(N) \propto 1/N^{2.2}$ developing in the saturation regime corresponding to high neutron exposure. %
Below, we show that this fact, also confirmed by experimental observations \citep{Ungar2021b}, has important implications for the interpretation of experimental observations of dislocation loops and for our understanding of the dynamics of microstructural evolution in the irradiated cladding. %

Our manuscript is structured by detailing our methods in \S \ref{sec:theory_methods} and presenting our results in \S \ref{sec:results} before summarising key conclusions in \S \ref{sec:conclusions}. %
The concept and relevant definitions of dislocation density are discussed in \S \ref{subsec:dis_dens}. %
Details of our experimental set-up and and an overview of the \textsc{cmwp} line profile analysis software are given in \S \ref{subsec:exp} followed by an explanation of our simulation method, its range of validity and our choice of settings in \S \ref{subsec:CRA}. %
We show how the simulated microstructure correlates to the dislocation density profile shown in Figure \ref{fig:CRA_XRD_comparison} in \S \ref{subsec:struct_distribution} in addition to characterising the evolution of the power law distribution of dislocation sizes. %
Finally, the stored energy associated with dislocation content that drives the changing microstructure as a function of dose is investigated in \S \ref{subsec:stored_energy}. %

\section{Theory and Methodology}
\label{sec:theory_methods}

\subsection{Dislocation density}
\label{subsec:dis_dens}

There are multiple ways to define the density of dislocation lines $\rho$ in a deformed or irradiated material. %
For all the measurements and computations in this study, $\rho$ is defined as a \textit{scalar} ratio of the total dislocation line length to volume $V$ containing the dislocations, namely \citep{Hull2011}  %
\begin{align}
	\rho := \frac{1}{V} \int_{\perp \in V}{ \left | \mathrm{d}\mathbf{l} \right |}.
 \label{eq:vol_dens}
\end{align}Here $\mathbf{l}$ is a position vector on a dislocation line such that $\mathrm{d}\mathbf{l} = \boldsymbol{\xi} \mathrm{d}s$ for unit tangent vector $\boldsymbol{\xi}$ and arc length $s$, and the integration is performed with respect to the arc length over all the dislocation lines $\perp$ in $V$. %
The choice of volume is somewhat arbitrary but, for a given resolution, is expected to reflect the average amount of dislocations present. %
For example, in a TEM micrograph this can be chosen to be a region contained in the image and in a molecular dynamics simulation one may use the entire simulation cell. %

Another possible definition is the areal density $\rho_A$ that measures the number of dislocation lines crossing an open surface as \citep{Hull2011} %
\begin{align}
	\rho_{A} := \frac{1}{A} \int_{\perp \in A}{\left | \mathrm{d}\mathbf{S} \cdot  \boldsymbol{\xi} \right |},
 \label{eq:area_dens}
\end{align}
where $\mathrm{d}\mathbf{S}$ is a vector area element of $A$ with direction normal to the surface and, similar to Equation \ref{eq:vol_dens}, there is a dependence on the choice of the surface. %
If all the dislocations are co-linear and perpendicular to the chosen surface, Equations \ref{eq:vol_dens} and \ref{eq:area_dens} produce the same value. %
Whilst for an arbitrary distribution of dislocations this is often not the case, generally both measures do not differ by much more than a factor of two \citep{Schoeck1962}. %

Needless to say, a dislocation is not solely defined by its tangent vector and it is noteworthy that neither Equation \ref{eq:vol_dens} nor Equation \ref{eq:area_dens} contain any information about the Burgers vector $\mathbf{b}$ of the dislocation. %
Significant physical quantities such as dislocation energy and the Peach-Koehler force both depend on $\mathbf{b}$. %
The Nye tensor $\boldsymbol{\alpha}$ is a \textit{tensorial} measure of dislocation density that is a linear function of lattice curvature \citep{Nye1953} and is also a function of position $\mathbf{x}$ such that \citep{Jones2016} %
\begin{align}
	\boldsymbol{\alpha}(\mathbf{x}) := \int_{\perp} {\delta\left(\mathbf{x} - \mathbf{l} \right) \mathbf{b}(\mathbf{l}) \otimes \mathrm{d}\mathbf{l} },
\label{eq:Nye_dens}
\end{align}
where $\otimes$ denotes the tensor product, integration is performed over all of the dislocation lines in the system, and $\delta \left( \mathbf{x} \right)$ is the Dirac delta distribution defined by the property
\begin{align}
    \int \mathrm{d}^3\mathbf{x}'\, \delta (\mathbf{x}' - \mathbf{x}) f(\mathbf{x}') = f(\mathbf{x}),
\end{align}
for an arbitrary well-behaved function $f(\mathbf{x} )$. %
Whilst full information about the dislocation content is contained in Equation \ref{eq:Nye_dens}, attempting to average $\boldsymbol{\alpha}$ over a volume $V$ can be problematic. %
Essentially, this stems from the fact that the integral of $\mathrm{d}\mathbf{l}$ along a dislocation segment contained in $V$ that starts and ends at $\mathbf{x}_{a}$ and $\mathbf{x}_{b}$ respectively is $\mathbf{x}_{b} - \mathbf{x}_{a}$. %
Thus, any information pertaining to curvature of a dislocation line is lost and closed paths in particular, \textit{i.e.}\ dislocation loops, provide no contribution to the volume average \citep{Arsenlis1999,Mandadapu2014}. %
The dislocations sections that integrate to zero are referred to as Statistically Stored Dislocations (SSD) and the surviving contributions are the Geometrically Necessary Dislocations (GND). %
Experimental techniques that infer GND content, such as micro-beam Laue measurements and high resolution electron back-scattered diffraction, implicitly make use of Equation \ref{eq:Nye_dens}  \citep{Das2018}. %
As it is difficult to characterise the entire population of dislocations using Equation \ref{eq:Nye_dens}, throughout this manuscript we have chosen to use the scalar measure given by $\rho$ in Equation \ref{eq:vol_dens} which is the same definition as that employed in our X-ray line profile analysis. %

\subsection{Experiment}
\label{subsec:exp}

Dislocation densities  were measured in four {\SI{3}{\mm}\,$\times$\,\SI{1}{\mm}\,$\times$\,\SI{0.5}{\mm}} Zircaloy-4 samples (composition Zr-0.17Fe-1.24Sn-0.10Cr) proton-irradiated to different doses. %
The samples possessed a recrystallised equiaxed microstructure with a low dislocation density of \SI{< 1e14}{\m^{-2}} and a characteristic `split-basal' texture due to processing, where the basal poles are aligned along the normal direction (ND), with a $\pm$30 degree tilt towards the transverse direction. %
Irradiation induced growth strains in similarly textured Zircaloys are known to saturate at doses below \SI{\sim 10}{dpa} at \SI{320}{\degreeCelsius} \citep{Adamson2019} and thus we may expect dislocation densities to display a similar pattern of evolution in these samples. %

The ND face of each sample was proton irradiated with \SI{2}{MeV} protons at \SI{350}{\degreeCelsius} at the University of Manchester's Dalton Cumbrian Facility, UK. %
The temperature of the samples during irradiation was monitored \textit{via} a thermal imaging camera in order to hold it within \SI{\pm 10}{\degreeCelsius} of the target temperature. %
Unlike neutrons, the Coulomb interaction between the protons and the target material results in the shallow penetration of protons into the material. %
The resulting radiation exposure, quantified by the dose and dose rate, varies significantly as a function of depth, with the dose rate being of the order of $10^{-5}$ dpa/s. %
The dose profile was calculated using the quick Kinchin-Pease setting in \textsc{srim} \citep{Ziegler2010} with the lattice binding energy and threshold displacement energy set to \SI{0}{eV} and \SI{40}{eV} respectively \citep{Stoller2013}. %
A typical dose \textit{vs}.\ depth profile in one of our Zircaloy-4 samples consists of a plateau region extending \SI{\sim 10}{\micro\m} from the surface where the dose and dose rate are approximately constant before sharply rising and falling to zero at a region corresponding to protons coming to rest in the material, called the Bragg peak and located at \SI{\sim 30}{\micro m}. %
The samples were irradiated such that the doses at 60\% of the Bragg peak depth from the surface, termed `nominal doses', were 0.1, 0.5, 1 and \SI{2}{dpa} respectively. %
Within the first \SI{30}{\um} of each sample from the irradiated surface, the calculated dose and dose rate vary from their nominal values by factors of 0.6 to 8.5 thus allowing us to measure data spanning over a wide range of irradiation exposures. %

\begin{figure}
\centering
    \begin{subfigure}{1\textwidth}
    \centering
    \includegraphics[width = 0.9\textwidth]{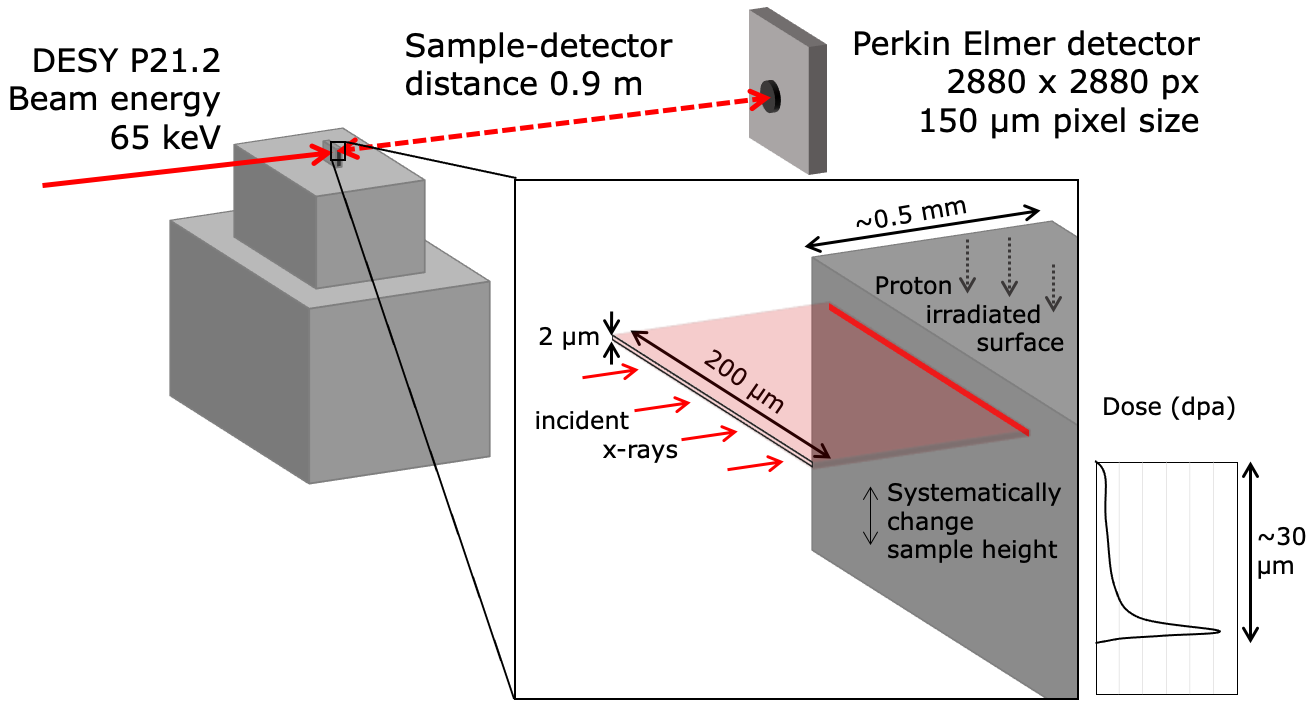}
    \caption{}
    \label{subfig:xray_setup}
    \end{subfigure} \\
    \begin{subfigure}{0.6\textwidth}
    \centering
    \includegraphics[width = 1\textwidth]{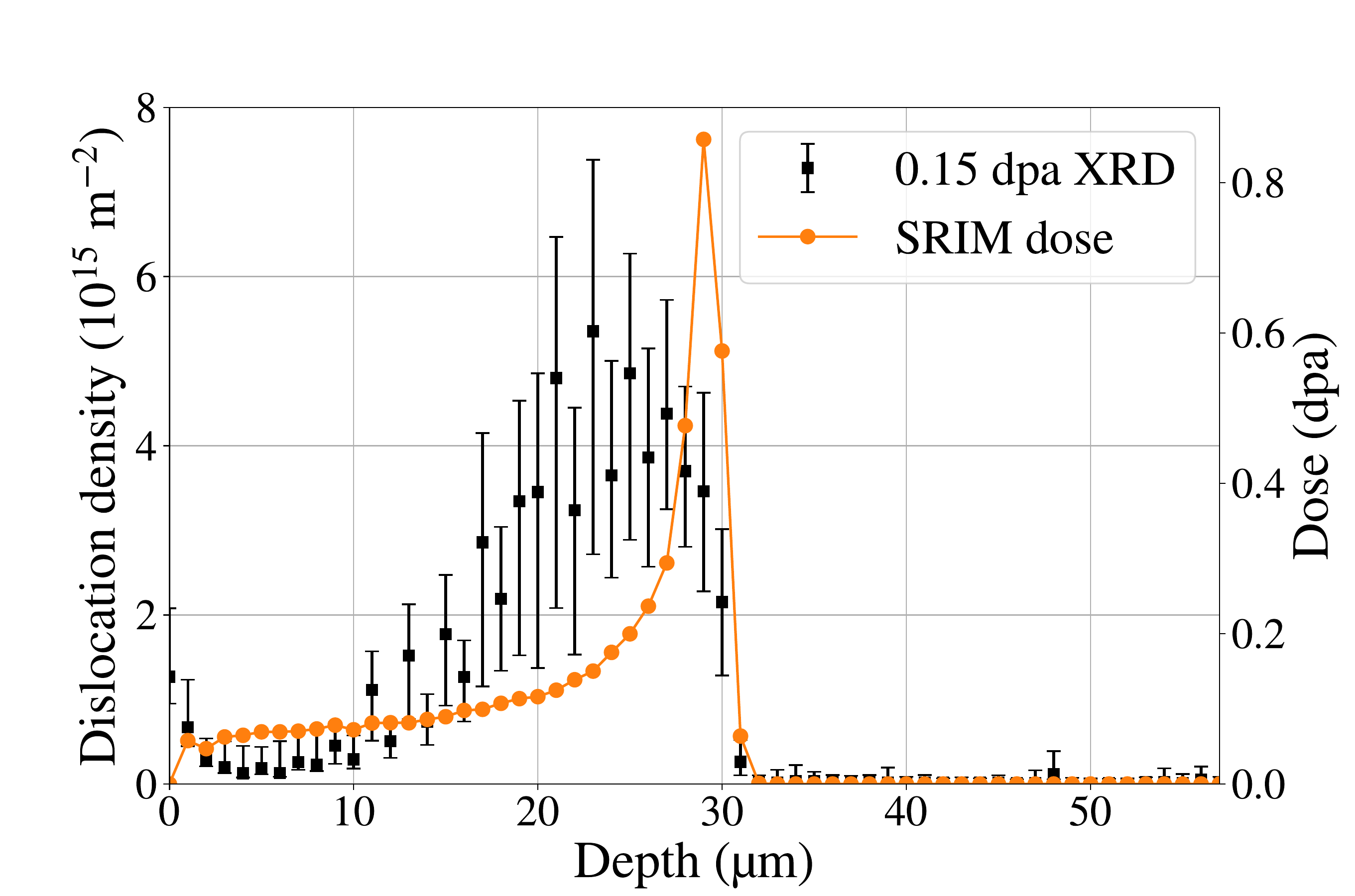}
    \caption{}
    \label{subfig:DD_SRIM}
    \end{subfigure}
    \caption{ (a) : Experimental geometry for measuring line profiles from Zircaloy-4 samples on the P21.2 beamline at the DESY synchrotron, Hamburg, Germany.  (b): Variation of dislocation density and dose as a function of depth for the nominal \SI{0.1}{dpa} sample.} 
    \label{fig:exp_figure}
\end{figure}

Using a small X-ray beam of \SI{2}{\um} $\times$ \SI{100}{\um} cross section, the samples were scanned in cross-section from the surface to a depth of \SI{50}{\um} within the sample at \SI{2}{\um} increments at the P21.2 beamline at the PETRA III synchrotron facility at DESY in Hamburg, Germany. %
The samples were translated perpendicular to the scanning direction by \SI{200}{\um} during each scan to improve grain statistics and reduce the spottiness of the pattern. %
The set-up of the diffraction experiment and sample geometry is shown in Figure \ref{fig:exp_figure}. %

Dislocation densities were extracted from the line profiles using the Convolutional Multiple Whole Profile (\textsc{cmwp}) software \citep{Ribarik2020}. %
The \textsc{cmwp} software models the line profile intensity $I(q)$, where $q$ is the wavevector magnitude, as a convolution of intensities arising from instrumental effects, size broadening, and, in particular, strain broadening due to dislocations. %
Instrumental effects were determined using a LaB$_6$ standard specimen, and the size broadening was determined assuming a log-normal size distribution of coherently scattering grains \citep{Ribarik2020}. %
Here, we outline the model underpinning the \textsc{cmwp} software, whereas for a broader context we refer an interested reader to reviews of the method \citep{Wilkens1970, Ungar1999,Ribarik2020}. %

In the theory of X-ray line profile analysis, the Fourier components of the broadened intensity peak profile corresponding to a reciprocal lattice vector $G$, denoted $I^D_{G}$, are related to the strain distribution by
\begin{align}
 \mathcal{F} \left\{ I^D_G(q) \right\}(L) = \exp \left[- 2 \pi^2 G^2 L^2 \langle \epsilon^2_{G,L} \rangle \right],
\end{align}
where $L$ is the Fourier variable and $\langle \epsilon^2_{G,L} \rangle$ is the mean-square strain. %
\cite{Wilkens1970} derived an expression for $\langle \epsilon^2_{G,L} \rangle$ by numerical methods arising from the `restrictedly random distribution' concept of dislocations. %
Dislocation lines are assumed to be parallel in the sub-areas of equal size $A$ perpendicular to the line direction. A number $N_{\perp}$ of dislocation lines with equal numbers of positive and negative Burgers vectors occupy random positions in a plane normal to the dislocation lines. %
The characteristic linear size of the sub-areas is chosen to be proportional to a parameter termed the effective outer cut-off radius $R_e$. %
The dislocation density $\rho$ is then defined by \cite{Wilkens1970} as the areal density of dislocations given by Equation \ref{eq:area_dens} that, as discussed in \S \ref{subsec:dis_dens}, is equivalent to the volume density of dislocations defined by Equation \ref{eq:vol_dens} for this specific  dislocation configuration. %
The dipole character of the distribution is determined by the arrangement parameter $M = R_e \sqrt{\rho}$ . %
Whilst $R_e$ is one of the fitting parameters in the \textsc{cmwp} software, thus affecting the value of $M$, this article is principally concerned with the determination of $\rho$ and thus we do not discuss the arrangement parameter further. %

An expression for the mean-square strain for a restricted random distribution of dislocations was derived by \cite{Wilkens1970} 
\begin{align}
\langle \epsilon^2_{G,L} \rangle = \frac{\rho C b^2}{4 \pi} f(\eta),
\end{align}
where $C$ is a parameter that is refined by the profile fitting \textsc{cmwp} algorithm, termed the dislocation contrast factor. The value of $C$ was evaluated for dislocations in $\alpha$-Zr by \cite{Balogh2016}. %
The Wilkens function $f(\eta)$, where $\eta = L/R_e$, has the following asymptotic forms in the limit of small and large $\eta$, respectively
\begin{align}
f(\eta) \sim 
    \begin{cases}
        \ln{\eta} &, \, \eta \rightarrow 0 \\
        \frac{1}{\eta} &, \, \eta \rightarrow \infty.
    \end{cases}
\end{align}
The numerically obtained formula for $f(\eta)$ may be found in \cite{Wilkens1970}. %

Although the Wilkens model is formally derived assuming that a dislocation configuration is composed of straight lines, the model is in fact able to mimic the statistical properties of distributions of curved dislocations \citep{Kamminga2000, Groma2004}.
When compared to transmission electron microscope (TEM) measurements of irradiated Zircaloy-2, the \textsc{cmwp} software was able accurately follow the dislocation density evolution as a function of dose \citep{Seymour2017}, and the \textsc{cmwp} approach has now become an accepted tool for determining dislocation densities \citep{Ungar2021a, Topping2018} as well as other microstructural features \citep{Ungar2021b} in irradiated Zircaloys. %
The \textsc{cmwp} software evaluates parameters describing effects of both the specimen size and dislocation broadening of diffraction intensity peaks by first employing a statistical Monte Carlo optimisation followed by the Marquardt-Levenberg non-linear least squares algorithm, see \cite{Ribarik2020}. %
All the peaks in the interval from \SI{0}{nm^{-1}} $< q <$ \SI{13}{nm^{-1}} were included in the fitting procedure and the uncertainty in $\rho$ was quantified according to the quality of fit as described by \cite{Ribarik2020}. %
The variation of dislocation density with depth calculated by \textsc{cmwp} is shown in Figure \ref{subfig:DD_SRIM} that, when mapped to the dose calculated by \textsc{srim} at a given depth, enables plotting the dislocation density as a function of dose as illustrated in Figure \ref{fig:CRA_XRD_comparison}. %

\subsection{Simulation}
\label{subsec:CRA}

The accumulation of defects and microstructural evolution as a function of dose was simulated using the Creation Relaxation Algorithm (CRA) \citep{DerletDudarev2020}. %
The CRA exploits the separation of timescales associated with relatively fast stress driven and comparatively slow thermally activated evolution of defect microstructure. %
This results in a simple algorithm where, starting with a perfect crystal structure of $\alpha$-Zr, the Frenkel pairs of defects are created at random and the microstructure is subsequently relaxed \textit{via} direct energy minimisation such that the system evolves purely through the action of internal stresses arising from the generation of defects. %
The dose is measured in units of `canonical displacement per atom' (cdpa) computed as the ratio of the total number of Frenkel pairs generated by the algorithm to the number of atoms in the system. 
CRA simulations assume that vacancies remain effectively immobile, in turn also immobilising the dislocation part of the microstructure \citep{Arakawa2020}, leaving the internal fluctuating stresses as the only remaining factor driving the migration and clustering of self-interstitial atom defects. %
In \citep{Warwick2021} we identified the approximate temperature and dose rate range where the simulation method retains its validity when applied to $\alpha$-Zr. %

The IIG strains observed in neutron irradiated zirconium alloys with initially low dislocation densities tend to saturate with increasing dose at temperatures less than \SI{\sim 300}{\degreeCelsius}, see \cite{Adamson2019}. %
At temperatures above \SI{\sim 300}{\degreeCelsius}, saturation persists over shorter intervals of dose before the strain magnitudes start increasing linearly as a part of the breakaway growth phenomenon \citep{Holt1988}. %
A significant change of pattern of thermal evolution has also been found above \SI{\sim 300}{\degreeCelsius} in proton irradiated Zircaloy-2 when samples irradiated to \SI{2}{dpa} were annealed for \SI{1}{\hour} at various temperatures \citep{Topping2018}. %
The X-ray line profile measurements performed by \cite{Topping2018} showed that the \textit{a}-loop density significantly decreased only at temperatures above \SI{\sim 300}{\degreeCelsius}. %

These data offer a valuable insight into the timescales on which thermally activated processes, including vacancy migration, drive the evolution of \textit{heavily} irradiated zirconium. %
Earlier \citep{Warwick2021}, noting that the rates of thermally activated processes follow the Arrhenius law \citep{Vineyard1957,Landauer1961,Allnatt1993}, we showed that the annealing experiment data imply that the characteristic activation energy $E_a$ for the processes primarily responsible for the observed thermally activated behaviour must be close to \SI{\sim 2}{eV}. %
Also, as described in \S \ref{subsec:exp}, the proton-irradiation defect production dose rate $\dot{\phi}$ at all depths in our experiments is high and close to  {$10^{-5}$ dpa s$^{-1}$}. %
Given this high dose rate, we can estimate an upper bound $\tilde T$ on the range of temperatures where the rate of migration of defects stimulated directly by irradiation is higher than the rate of thermally activated migration of defects. For a given activation energy $E_a$, using the dose rate model by  \cite{Nordlund2018}, we find that the two rates are comparable if 
\begin{align}
    \dot{\phi} \left ( {2E_d\over E_a}\right) \approx  \nu \exp{ \left(- \frac{E_a}{k_B \tilde T} \right) },
\label{eq:arrhenius}
\end{align}
where $E_d$ is the threshold displacement energy required for forming a defect, the attempt frequency is ${\nu \approx \omega_D / 2\pi}=5.84\times 10^{12}$ s$^{-1}$, given the Debye frequency ${\omega_D = 3.67 \times 10^{13} \, \mathrm{s}^{-1}}$ \citep{Zarestky1979} and $k_B=0.861 \times 10^{-4}$ eV/K is the Boltzmann constant. %
Taking $E_a=2$ eV and $E_d=40$ eV, and solving equation (\ref{eq:arrhenius}) for $\tilde T$, we find $\tilde T =625$ K\SI{\approx  350}{\degreeCelsius}. Below this temperature, the eigenrate of thermal relaxation of microstructure is lower than the rate at which defects are driven by irradiation. Hence, at temperatures below $\tilde T$, the  defect structures generated by irradiation evolve predominantly through fast athermal stress relaxation \citep{DerletDudarev2020}.

The above esitmate for $\tilde T$\SI{\sim 350}{\degreeCelsius} is close to the temperature at which \cite{Topping2018} observed the occurrence of a significant change in the thermal response of microstructure during annealing. Notably, the change of pattern of breakaway growth at \SI{\sim 300}{\degreeCelsius} noted by \cite{Holt1988} was observed at significantly lower dose rates than those characterising our proton irradiation experiments. Hence, the above temperature must reflect the fundamental scale of activation energies associated with microstructural evolution of Zircaloys under irradiation.

The migration energy of individual vacancies in pure elemental $\alpha$-Zr \citep{Varvenne2014} of \SI{\sim 0.5}{\eV} is too low to account for the observed behaviour, and while the thermal diffusion of vacancies and other point defects affect microstructural evolution, reducing the overall concentration of defects noted in our analysis, experimental observations indicate the presence of a rate-limiting process with a higher activation energy that stabilises the observed dense defect microstructures at temperatures as high as \SI{300}{\degreeCelsius}. 

High activation energies are known to be associated with the formation of immobile vacancy-impurity clusters involving carbon or nitrogen \citep{Fu2008,Terentyev2014,Apostolopoulos2022}. In bcc iron, the {\it effective} migration energy of vacancies is defined by the energy of dissociation of a cluster involving a vacancy and a carbon dimer \citep{Paxton2014}, and this dissociation energy can be as high as 2.22 eV \citep{Kabir2010}, far higher than the activation energy of 0.55 eV characterising vacancy migration in pure elemental Fe \citep{Fu2005}. Given that the characteristic formation and migration energies of defects in Zr and Fe are nearly the same \citep{Dudarev2013}, the high effective activation energy of the order of 2 eV seen in experiments on Zr likely result from the impurity effect similar to that found in Fe. As noted by \cite{Kabir2010}, at relatively low temperatures the vacancy-carbon dimer complexes are immobile, making the dissociation temperature of these complexes one of the key parameters determining the response of a material to radiation exposure.

CRA simulations \citep{DerletDudarev2020} or the simulations involving the production of defects by successive collision cascade events \citep{Mason2021,Granberg2023,Boleininger2023} do not imply the absence of mobility of defects. Self-interstitial atom defects exhibit the non-Arrhenius mobility \citep{Dudarev2008} and their motion is strongly affected by elastic strain fields \citep{Dudarev2010}, resulting in the rapid clustering of these defects into interstitial dislocation loops and, subsequently, into a dense entangled  network of dislocations  \citep{DerletDudarev2020,Boleininger2022}. The latter forms spontaneously at doses above approximately \SI{0.3}{dpa} \citep{Mason2020}. The clustering of self-interstitial defects into dislocation loops and dislocations stems from the fact that this is a highly energetically favourable process, releasing up to $E^f_{SIA}\approx 3$ eV per self-interstitial coalescence event \citep{Domain2005,Dudarev2013}. 

The fact that it is the SIA formation energy, fundamentally related to the strong elastic interaction between the self-interstitial defects, that drives the evolution of microstructure at relatively low tempeatures rather than the diffusion of self-interstitial  {\it per se}, is confirmed by finite-temperature simulations by \cite{Chartier2019}. The simulations were performed at 300K and hence included the thermal diffusion of self-interstitial atom defects, but still exhibited the same pattern of evolution as that predicted by the CRA simulations \citep{DerletDudarev2020}. This is confirmed by experimental observations by \cite{Wang2022} showing the trends similar to those found in simulations, even though in tungsten the diffusion of self-interstitial defects occurs at temperatures as low as 27 K \citep{LandoltBornstein1991,Ma2019}. 

The above analysis of {\it ab initio} data and experimental information shows that the temperature interval over which the dynamics of microstructural evolution changes from the low-temperature mode dominated by  microscopic stress fluctuations \citep{DerletDudarev2020} to the high-temperature mode dominated by the Arrhenius thermally activated diffusion \citep{Allnatt1993}, in zirconium alloys spans approximately from \SI{300}{\degreeCelsius} to \SI{400}{\degreeCelsius}, as illustrated particularly well by Fig. 8 from \cite{Topping2018}. The qualitative picture of microstructural evolution of zirconium irradiated by high-energy protons can now be summarised as follows. Proton irradiation produces relatively low energy recoils, generating defects in the form of Frenkel pairs or small defect clusters \citep{Boleininger2023}. Self-interstitial atom defects coalesce into dislocation loops and a dislocation network, whereas vacancies either diffuse and recombine with the interstitial dislocation loops or extended dislocation structures, or form immobile vacancy-impurity clusters \citep{Kabir2010}. These vacancy-impurity clusters immobilise and stabilise the dislocation microstructure \citep{Arakawa2020}, but dissociate in the temperature interval from \SI{300}{\degreeCelsius} to \SI{400}{\degreeCelsius}. Over this temperature interval, the mode of microstructural evolution changes from that dominated by stress fluctuations and coalescence of self-interstitial defects to the mode dominated by vacancy diffusion. Over the same interval of temperatures, the IIG changes from a mode exhibiting saturation to that of runaway growth. Our observations exhibit the formation of a dense dislocation network, suggesting that the experimental conditions correspond to the low-temperature rather than the high-temperature mode of microstructural evolution. The selection of a simulation approach below reflects and recognises this fact. An algorithm for modelling microstructural evolution at higher temperatures has to include the treatment of microscopic stress fluctuations as well as diffusion and interaction of vacancies and impurities. The development of such an algorithm remains a challenge for future studies. 

The CRA was implemented in the molecular dynamics program \textsc{lammps} \citep{Plimpton1995} \footnote{\url{https://lammps.sandia.gov}, 3 Mar and 29 Oct 2020 stable builds}. %
For this study, unless stated otherwise, we present results averaged across all three Embedded Atom Method (EAM) potentials developed in Ref.~\citep{Mendelev2007}, as is the case in Figure \ref{fig:CRA_XRD_comparison}. %
Whilst there are variations between the potentials with respect to their predicted formation energies and elastic properties of self-interstitials and vacancies \citep{Mendelev2007, Varvenne2014, Varvenne2017}, we have found that all three potentials qualitatively produce the same macroscopic dimensional changes and microstructural evolution under the CRA \citep{Warwick2021}. %
Furthermore, a similar study also employed the CRA on Zr and predicted the same trends with a different potential \citep{Tian2021b}. %

Our simulations employ periodic boundary conditions and supercells containing {$\sim$\,2M} and {$\sim$\,10M} atoms, with the cell edges parallel to the $[2 \bar1 \bar1 0]$, $[\bar1 2 \bar1 0]$ and $[0 0 0 1]$ directions. %
Energy minimisation was performed using a combination of the conjugate gradient and FIRE algorithms \citep{Bitzek2006} such that the relaxed force on any atom was smaller than \SI{1}{meV \text{\AA}^{-1}}. %
In the interest of computational efficiency, the simulation cell shape and size was kept fixed during relaxation. %
Whilst these boundary conditions result in a macroscopic internal stress of \SI{\sim 1}{GPa}, the dimensional changes that would occur if the cell shape relaxed may nevertheless be accurately computed using linear elasticity theory; furthermore, the microstructure resulting from relaxing under zero pressure is similar to that under zero strain \citep{Warwick2021, Tian2021b}. %
Dislocations were identified directly from atomic positions using the Dislocation eXtraction Algorithm (DXA) \citep{Stukowski2012}. %
This is achieved by assigning crystal structure types to each atom using common neighbour analysis \citep{Faken1994}. %
Given the crystal structure of the hcp reference crystal, Burgers circuits are drawn around regions containing atoms assigned to non-hcp crystal structure in order to compute Burgers vectors and dislocation lines. %
Dislocation densities are computed according to Equation \ref{eq:vol_dens}. %

In order to enable a closer comparison between our simulations and line profile analysis experiments, we simulated the intensity profile of powder diffraction patterns for all of the CRA microstructures using the Debye equation \citep{Debye1915} where the intensity of scattered X-rays is proportional to %
\begin{align}
    I(q) = \sum_{i,j}{ \mathrm{sinc} \left( 2 \pi q r_{ij} \right) }, 
\label{eq:debye}
\end{align}
for wavevector magnitude $q$, and the sum runs over all the pairs of atoms positioned at $\mathbf{r}_i$ and $\mathbf{r}_j$ separated by distance ${r_{ij} = |\mathbf{r}_i - \mathbf{r}_j|}$. %
Furthermore, in Equation \ref{eq:debye} it is assumed that the atomic form factor is the same for all atoms in the system. %
The time required for computing all the pairwise distances $r_{ij}$ for a system of $N$ atoms scales unfavourably as $N^2$ and thus we parallelised the task. %
The line profile was calculated without using periodic boundary conditions and thus the powder was treated as if it were composed of randomly oriented nano-grains as large as the simulation box. %
$I(q)$ was computed over the domain \SI{3}{nm^{-1}} $ \leq q \leq $ \SI{13}{nm^{-1}} spanning all peaks up to the $\{ 2 2 \bar{4} 0 \}$ reflections and the wavenumbers were sampled every \SI{2e-3}{nm^{-1}}. %

Data were visualised and processed with the \textsc{ovito} \citep{Stukowski2010b} and \textsc{paraview} \citep{AHRENS2005} software packages. %
For more details, please refer to our recent publication \citep{Warwick2021}. %

\section{Results and Discussion}
\label{sec:results}

\subsection{Dislocation structure and distribution}
\label{subsec:struct_distribution}
\begin{figure}
\centering
    \begin{subfigure}{.45\textwidth}
        \centering        \includegraphics[width=0.95\linewidth]{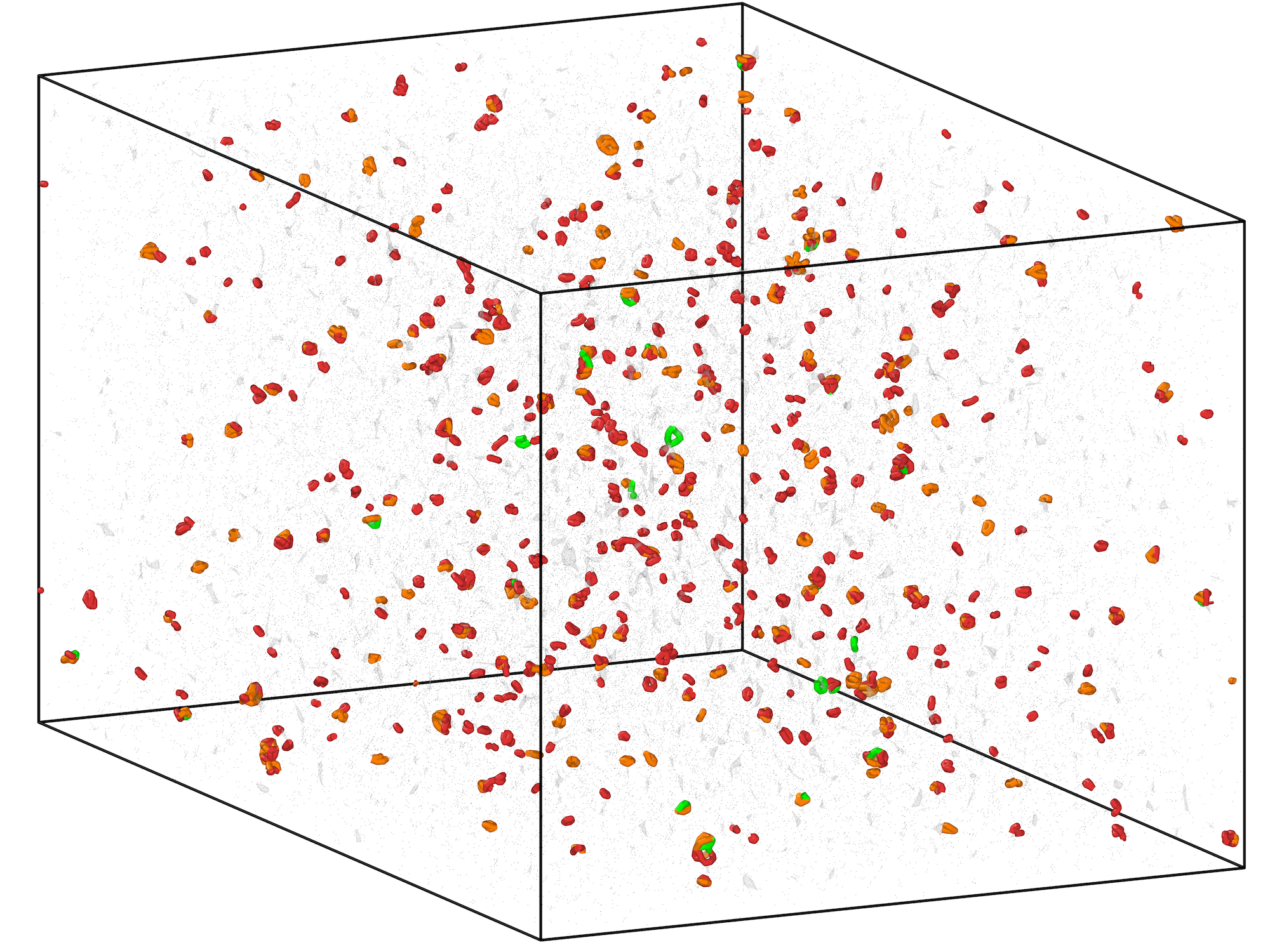}
        \caption{}
        \label{subfig:0.01cdpa}
    \end{subfigure}
    \begin{subfigure}{.45\textwidth}
        \centering
        \includegraphics[width=0.95\linewidth]{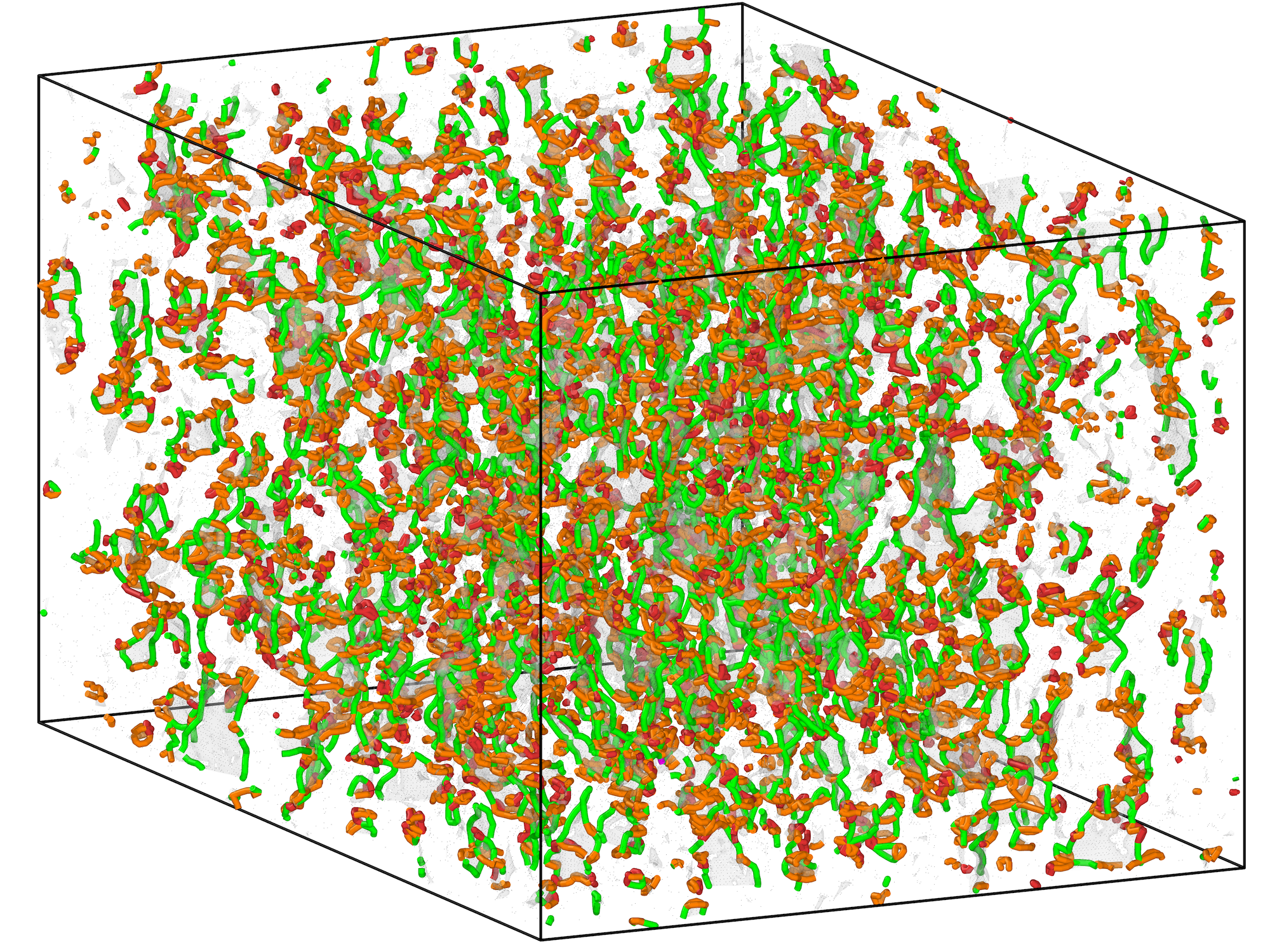}
        \caption{}
        \label{subfig:0.17cdpa}
    \end{subfigure} \\ 
    \begin{subfigure}{.45\textwidth}
        \centering
        \includegraphics[width=0.95\textwidth]{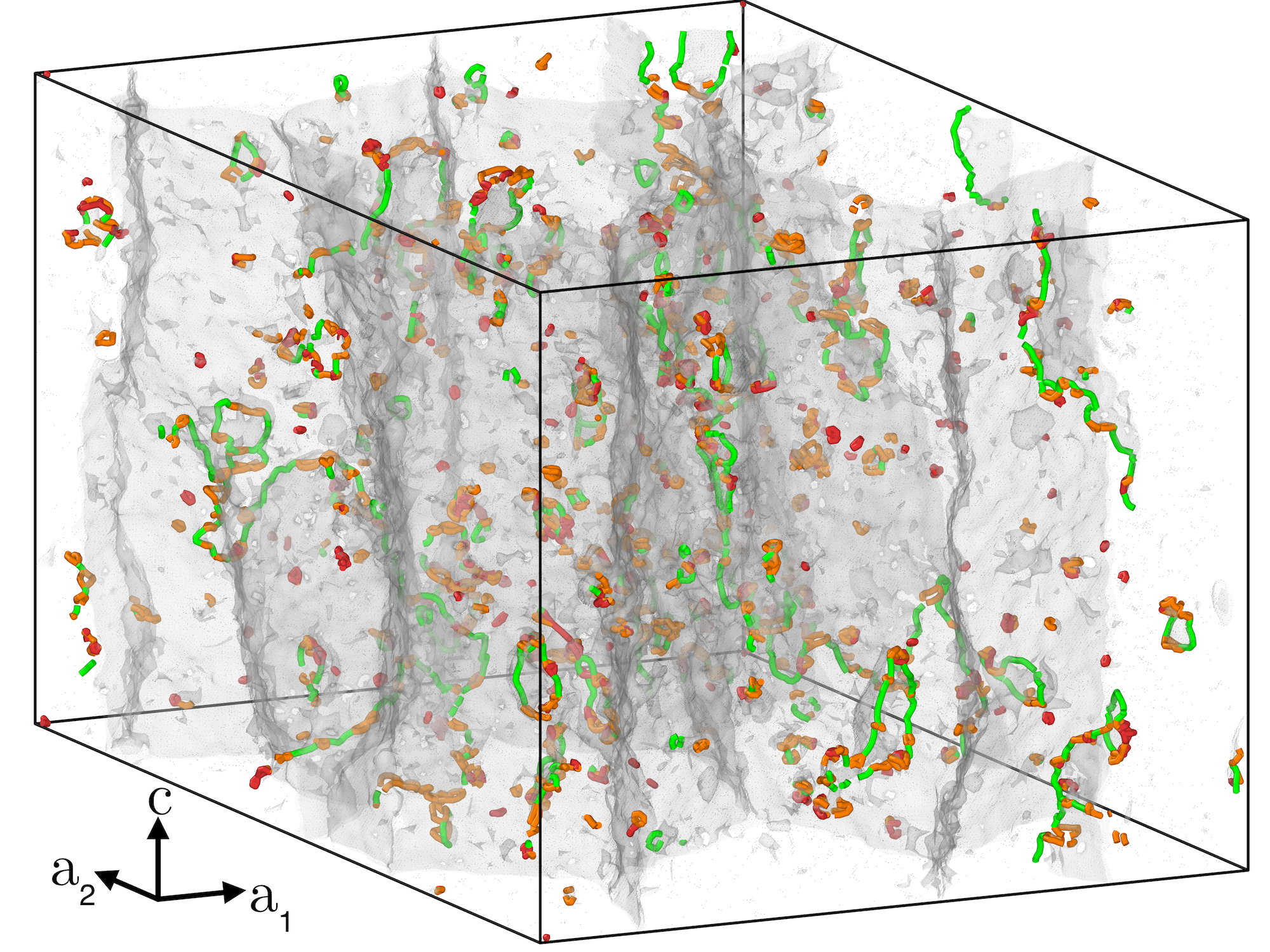}
        \caption{}
        \label{subfig:2.00cdpa}
    \end{subfigure}
    \caption{Dislocation microstructures calculated by the CRA simulations using the MA1 potential with 10M atoms. Green, orange and red dislocations indicate $\langle 2 \bar 1 \bar 1 0 \rangle , \langle 1 \bar 1 0 0 \rangle$ and unusual Burgers vectors, respectively. Dots are self-interstitial atoms identified with the Wigner-Seitz analysis and boundaries of interstitial clusters are rendered as transparent surfaces. \ref{subfig:0.01cdpa}: Self interstitial point defects and small dislocation loops at very low dose 0.02 cdpa. \ref{subfig:0.17cdpa} : Dense dislocation network at low dose 0.13 cdpa. \ref{subfig:2.00cdpa} : Extended dislocation network at high dose 2.00 cdpa with percolating interstitial cluster.}
    \label{figs:dislocation_renders}
\end{figure}

The peak and saturation of dislocation density shown in Figure \ref{fig:CRA_XRD_comparison} can be readily understood by inspecting the spatial configuration of dislocations generated by our simulations. %
The nature of defects evolving through internal stresses in the CRA results in dislocations being almost exclusively formed by the agglomeration of self-interstitial atoms whilst vacancies remain immobile and generally form small clusters containing $\mathcal{O} (10)$ vacancies that are not large enough to relax into dislocation loops. %
We do find a small number of much larger clusters containing $\mathcal{O}(100)$ vacancies. 
However, these are fundamentally interstitial in origin as the large vacancy clusters are nothing but vacant spaces in the crystallographic planes formed by the self-interstitial defects, see \cite{Mason2020,Boleininger2023}.
The renders shown in Figure \ref{figs:dislocation_renders} suggest that the dislocation structure evolves in three stages. %
At low dose (Figure \ref{subfig:0.01cdpa}), small loops form before coalescing into a dislocation network (Figure \ref{subfig:0.17cdpa}) at which point the dislocation density saturates. %
At high doses (Figure \ref{subfig:2.00cdpa}), full interstitial-type atomic planes form and the dislocation network fragments into loops, resulting in a drop in the dislocation density. %
The visualised microstructures were rendered from simulations employing the MA1 potential. %
When comparing the interatomic potentials, we discovered that the MA2 and MA3 potentials produce large populations of twinned regions \citep{Warwick2021}. %
It seems likely that these are artefacts of the potentials since such defects are not commonly observed in experiment. %
As was noted by \citet{Warwick2021}, for the MA3 potential in particular, a large proportion of dislocations coalesce into these twinned regions whose volume fraction also features a transient peak followed by saturation. %
The twinned regions are composed of dense arrays of dislocations and thus the pattern of evolution of dislocation structures and saturation in density is common to all three potentials. %
Thus, the microstructures derived from MA1 simulations are presented in order to avoid needlessly complicating our discussion. %

\begin{figure}
    \centering
    \includegraphics[width = 0.5\textwidth]{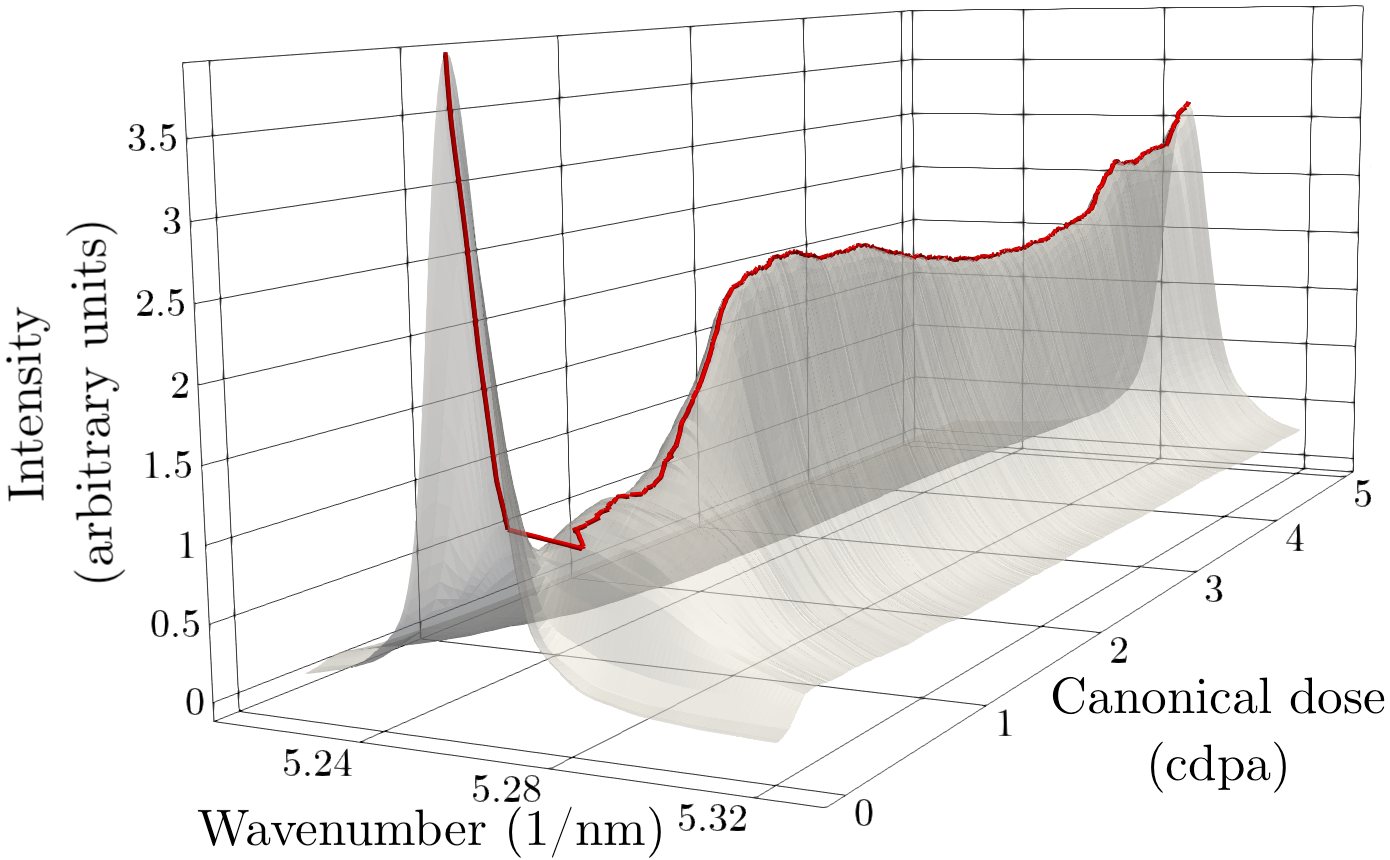}
    \caption{Simulated $\{1\bar{1}02\}$ X-ray diffraction peak profile as a function of dose computed from the MA1 microstructures. The variation in the intensity of the peak and its centre shift is highlighted by the red path. Application of the \textsc{cmwp} software to our simulated line profiles also shows that the dislocation density saturates as a function of dose, approaching a limiting value of $\sim 10^{15}$ m$^{-2}$.}
    \label{fig:waterfall_plot}
\end{figure}

CRA simulations tend to overestimate the defect content in the high dose limit \citep{Boleininger2023}, however the qualitative trends observed under a variety of conditions are predicted accurately \citep{Mason2020,Mason2021,Warwick2021}. %
This overestimation arises from the lack of re-crystallisation induced by collision cascades. %
Producing the primary knock on atoms with kinetic energies close to the threshold displacement energy results in defect densities very close to those predicted by the CRA whilst much larger recoil energies result in defect densities lower by approximately a factor of approximately 10, see \cite{Boleininger2023}.  
Thus, in agreement with the analysis by \cite{Mason2020,Mason2021,Boleininger2023} where the defect content was independently assessed using the experimentally measured deuterium retention, we have scaled down the dislocation density in Figure \ref{fig:CRA_XRD_comparison} by a factor of 10 to enable a direct comparison of our experimental and simulation data. %
After applying this scaling it can be seen that the CRA indeed predicts a qualitatively accurate variation dislocation density profile. %
The occurrence of a transient peak of dislocation density at a moderate dose was also noted earlier in simulations of iron and tungsten \citep{Chartier2019,DerletDudarev2020}. 

In order to enable a closer comparison with experiment we simulated a powder diffraction line profile for all of our CRA microstructures as described in \S \ref{subsec:CRA}.
Figure \ref{fig:waterfall_plot} illustrates the evolution of the $\{1\bar{1}02\}$ diffraction peak profile as a function of dose, reflecting the eventual saturation of the microstructure in the peak profile seen in the limit of high dose. %
We observe that in the transient regime, the peak intensity drops before rising and settling at a steady value. %
Furthermore, the peak broadens at high dose and shifts to higher scattering angles, indicating lattice compression. %
The formation of extra atomic planes due to the coalescence of dislocation loops does not result in the volumetric expansion but instead causes lattice compression because the simulations are performed at constant cell shape and size. %
Under zero pressure boundary conditions, we expect the peak centre to shift to lower wave-numbers instead. %

Saturation of the peak profile has been quantified by extracting the dislocation density from the data using the \textsc{cmwp} software. %
\textsc{cmwp} is known to infer dislocation densities that are larger than those determined from TEM images and this has been attributed to the ability of X-ray line profile analysis to resolve small loops in power law distributed dislocation loops \citep{Ungar2021a, Ungar2021b}. %
Interestingly, we find that the dislocation densities computed by \textsc{cmwp} and DXA nevertheless differ by approximately an order of magnitude although the character of variation of the observed and simulated dislocation densities as functions of dose are the same. %
We note that the difference between the \textsc{cmwp} software and DXA results brings attention to an important question: at what size is a dislocation loop too small to be counted as a dislocation object? %

Usually, dislocations are considered to be the sources of long-range strain fluctuations responsible for the X-ray peak line profile broadening detected in irradiated materials \citep{Wilkens1970}. %
On the other hand, small dislocation loops produce shorter-range strains and in the far-field limit they are equivalent to point defects, with the associated strain fields resulting in the Huang diffuse scattering, producing a relatively uniform increase in the scattered intensity in X-ray diffraction patterns \citep{Simmons1958}. %
Furthermore, when a dislocation loop is so small that the loop diameter is comparable to the core width of a dislocation then the loop is mostly comprised of core atoms and its structure can no longer be described using conventional elasticity \citep{Boleininger2018,Boleininger2019}. %
Thus, one would expect the resulting strains to be unlike those associated with linear elastic fields of dislocation loops and the core effects to be significant. %
Modelling the strain broadening effects associated with small dislocation loops requires further analysis and we defer it to a future study. %
A large proportion of the dislocation objects found in the simulated microstructures are so small that they should be treated as point defects by CMWP and would also not be easily detected in TEM images \citep{Zhou2006}. %
Determining the nature of such defects could be highly relevant to determining mechanisms that cause the complex high dose phenomena such as breakaway growth. %

\begin{figure}
\centering
    \begin{subfigure}{.45\textwidth}
    \centering
    \includegraphics[width = 0.9\textwidth]{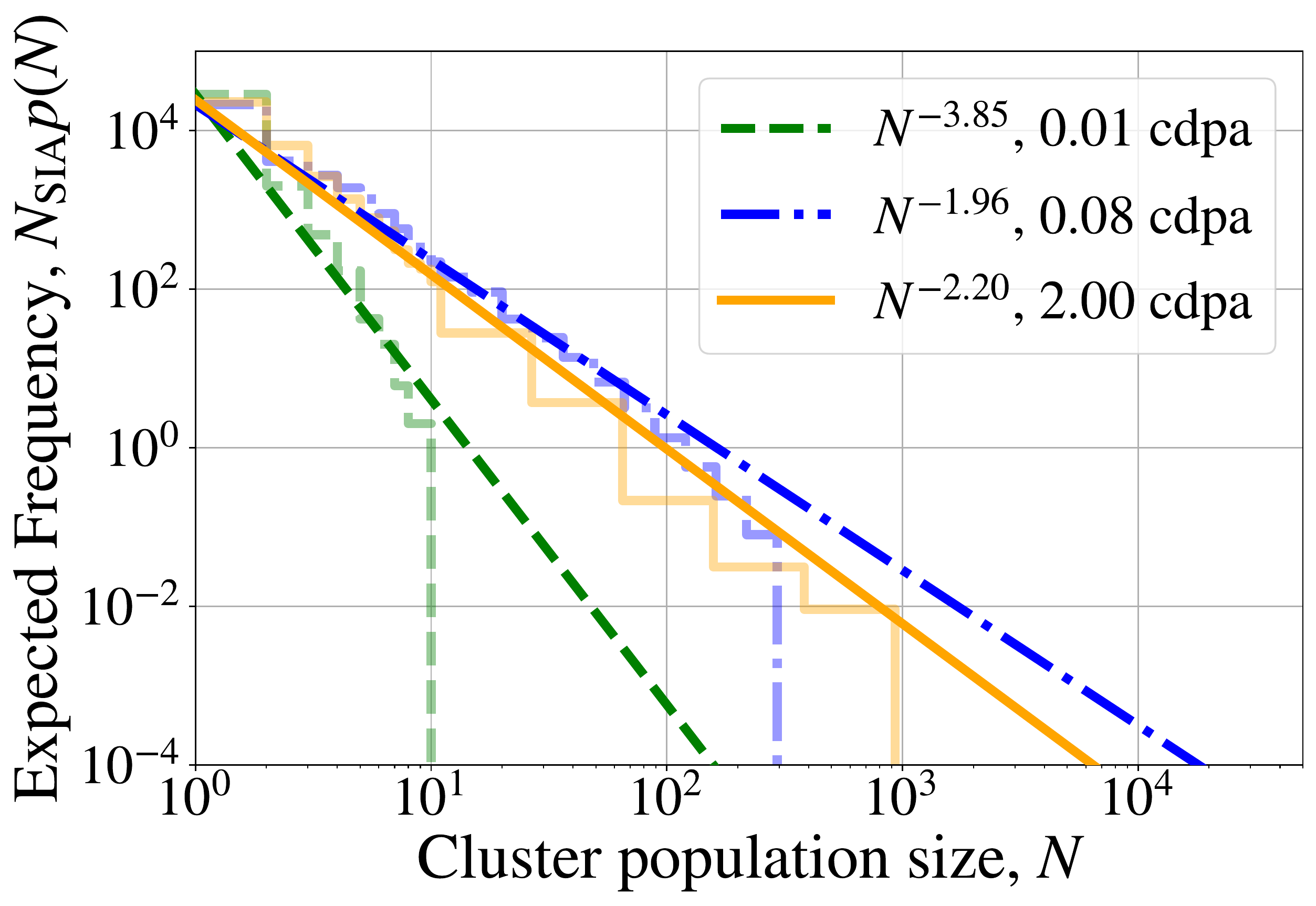}
    \caption{}
    \label{subfig:SIA_cluster_distribution}
    \end{subfigure}
    \begin{subfigure}{.49\textwidth}
    \vspace*{-0.45cm}
    \centering
    \includegraphics[width = 0.9\textwidth]{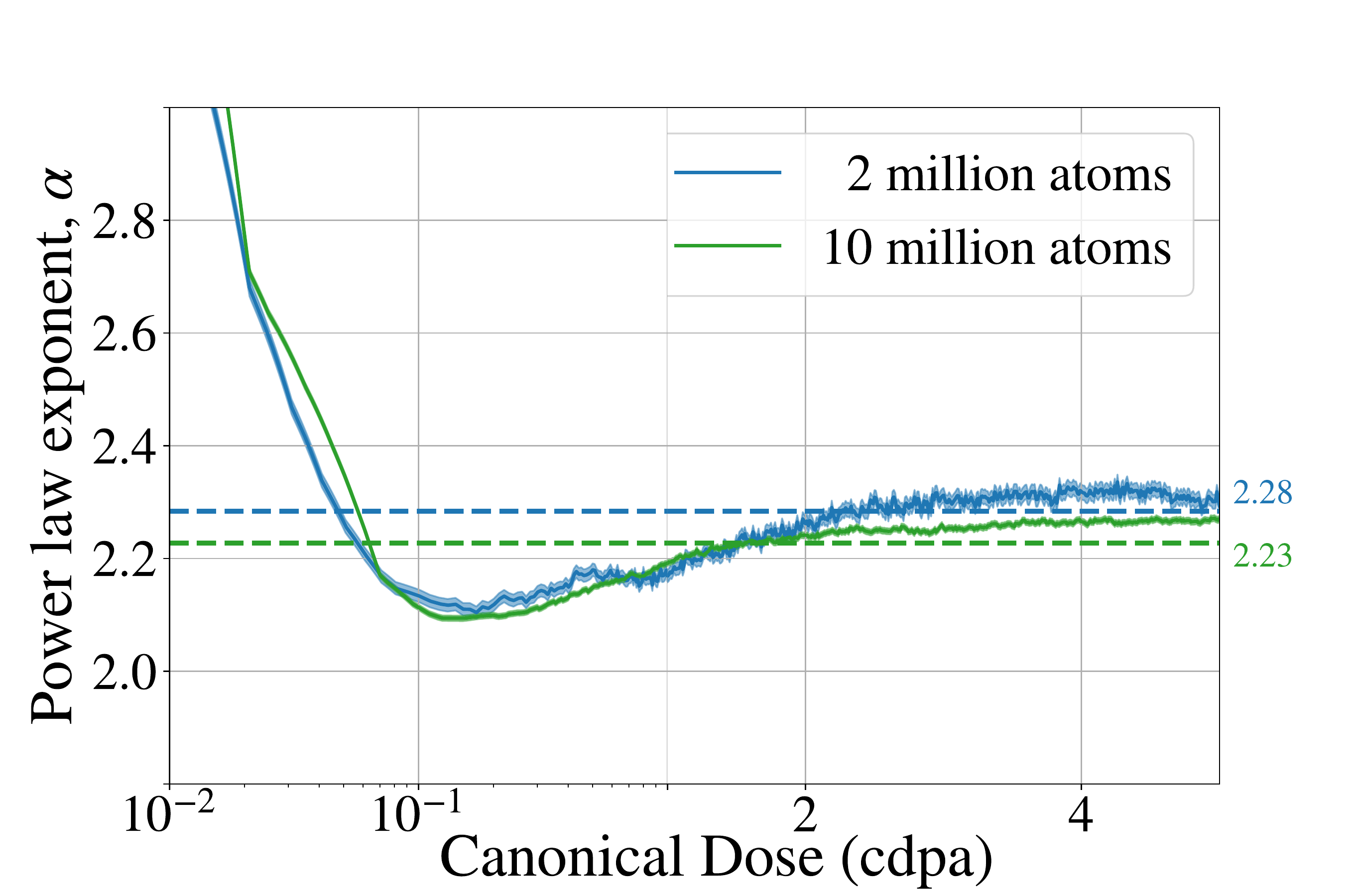}
    \caption{}
    \label{subfig:power_law_exponent_dose}
    \end{subfigure}
    \caption{ \ref{subfig:SIA_cluster_distribution} : Histogram of self-interstitial cluster population sizes detected in MA1 microstructures containing 10M atoms corresponding to Figures \ref{figs:dislocation_renders}. The population sizes of clusters, $N$, follow a power law distribution. $N_{\mathrm{SIA}}$ denotes the total number of self-interstitial atoms detected by the Wigner-Seitz algorithm. The data are presented on a log-log scale. Lines of best fit were calculated using the maximum likelihood estimates (MLE) for the exponent in a power law distribution stated in Eq.\ \ref{eq:powlaw}. \ref{subfig:power_law_exponent_dose}: Evolution of the MLEs for the power law exponent $\alpha$ as a function of dose averaged over all three interatomic MA potentials. Shaded regions represent confidence intervals of the MLE. Data for the doses lower than \SI{1}{cdpa} are presented on a log-linear scale whereas in the saturation regime \SI{>1}{cdpa} a linear-linear scale is employed. $\alpha$ was averaged over doses larger than \SI{1}{cdpa} to arrive at the value $\langle \alpha \rangle = 2.28 \pm 0.01$ indicated by the horizontal dashed blue line for a 2M cell, and $\langle \alpha \rangle = 2.23 \pm 0.01$ (dashed green line) for a 10M cell.} 
    \label{fig:SIA_cluster_distribution}
\end{figure}

Size distributions of dislocation loops are often investigated in experiment \citep{Yi2015} and so for the purposes of comparison and characterising the spatial arrangement of dislocations we examined the distribution of defect cluster - dislocation loop sizes. %
As the dislocations in this simulation are mostly of interstitial type, we can gain insight into the statistics of dislocation structures by examining the statistics of interstitial defect clusters. %
Figure \ref{subfig:SIA_cluster_distribution} shows the frequency distribution of clusters containing $N$ interstitials as calculated by \textsc{ovito} for the doses corresponding to the renders in Figure \ref{figs:dislocation_renders}. %
For clusters with population sizes $N < 10$, the bin widths of the histogram are equal to 1 whereas at larger cluster population sizes the bin widths increase logarithmically \citep{Milojevic2010}. %
The raw data for the cluster population size frequencies were averaged over the bin widths to produce the step chart shown. %
Visually, the histograms appear to follow a straight line on a log-log scale, suggesting a power law distribution. %
Furthermore, simulations employing the CRA have shown evidence for self-organised critical behaviour \citep{DerletDudarev2020} and thus it is reasonable to test the cluster population size power law distribution hypothesis such that the probability mass function for clusters containing $N$ interstitials is given by %
\begin{equation}
    p(N) = \frac{1}{\zeta(\alpha) N^\alpha}
    \label{eq:powlaw}
\end{equation}
where the Riemann zeta function \citep{Abramovitz} is defined as $\zeta(\alpha) = \sum_{k=1}^{\infty} 1 / k^\alpha $ for exponent $\alpha > 1$. %

We may define an exponent that best fits the data shown in Figure \ref{subfig:SIA_cluster_distribution} \textit{via} a maximum likelihood estimation. %
The likelihood function
\begin{align}
\mathcal{L}\left(\{N_i | i \in [1..N_{\mathrm{tot}}] \} | p(N | \alpha) \right) = \prod_{i=1}^{N_{\mathrm{tot}}} p(N_i | \alpha)
\label{eq:likelihood}
\end{align}
returns the probability of observing the $N_{\mathrm{tot}}$ measured data points $\{N_i | i \in [1..N_{\mathrm{tot}}] \}$ if they were produced from a distribution $p(N | \alpha)$ of given parameter(s) $\alpha$. %
The MLE is produced by determining the value of $\alpha_{\mathrm{MLE}}$ maximising $\ln\mathcal{L}$ such that:
\begin{align}
\left.\frac{\partial \ln\mathcal{L}}{\partial \alpha}\right\rvert_{\alpha = \alpha_{\mathrm{MLE}}} = 0.
\label{eq:likelihood}
\end{align}
Equation \ref{eq:likelihood} was solved numerically using the python package \textsc{powerlaw} \citep{Alstott2014} and as shown in Figure \ref{subfig:power_law_exponent_dose}, the MLEs and associated standard error for $\alpha$ exhibit transients over a range of doses corresponding to the formation of a dislocation network by the coalescence of loops. %
The minimum of this transient appears to be correlated with the peak in dislocation density before saturating to a constant value at high doses. %
The twinned regions that emerge when employing the MA2 and MA3 potentials, as described at the beginning of \S \ref{subsec:struct_distribution}, are also interstitial defect clusters. %
Therefore they are included in this statistical analysis and we find the same trend and remarkably similar values of $\alpha$ across all three MA potentials. %
Averaging across the potentials for doses larger then \SI{1}{cdpa} in the saturation regime, we find the MLE for exponent $\alpha = 2.28 \pm 0.01$ for a 2M simulation cell and $\alpha = 2.23 \pm 0.01$ for a 10M simulation cell. %
These values are close but slightly higher than the exponents found in simulations and observations of collision cascades in tungsten in the limit of low dose \citep{Sand2013,Yi2015}. %

\begin{figure}
    \centering
    \includegraphics[width = 0.5\textwidth]{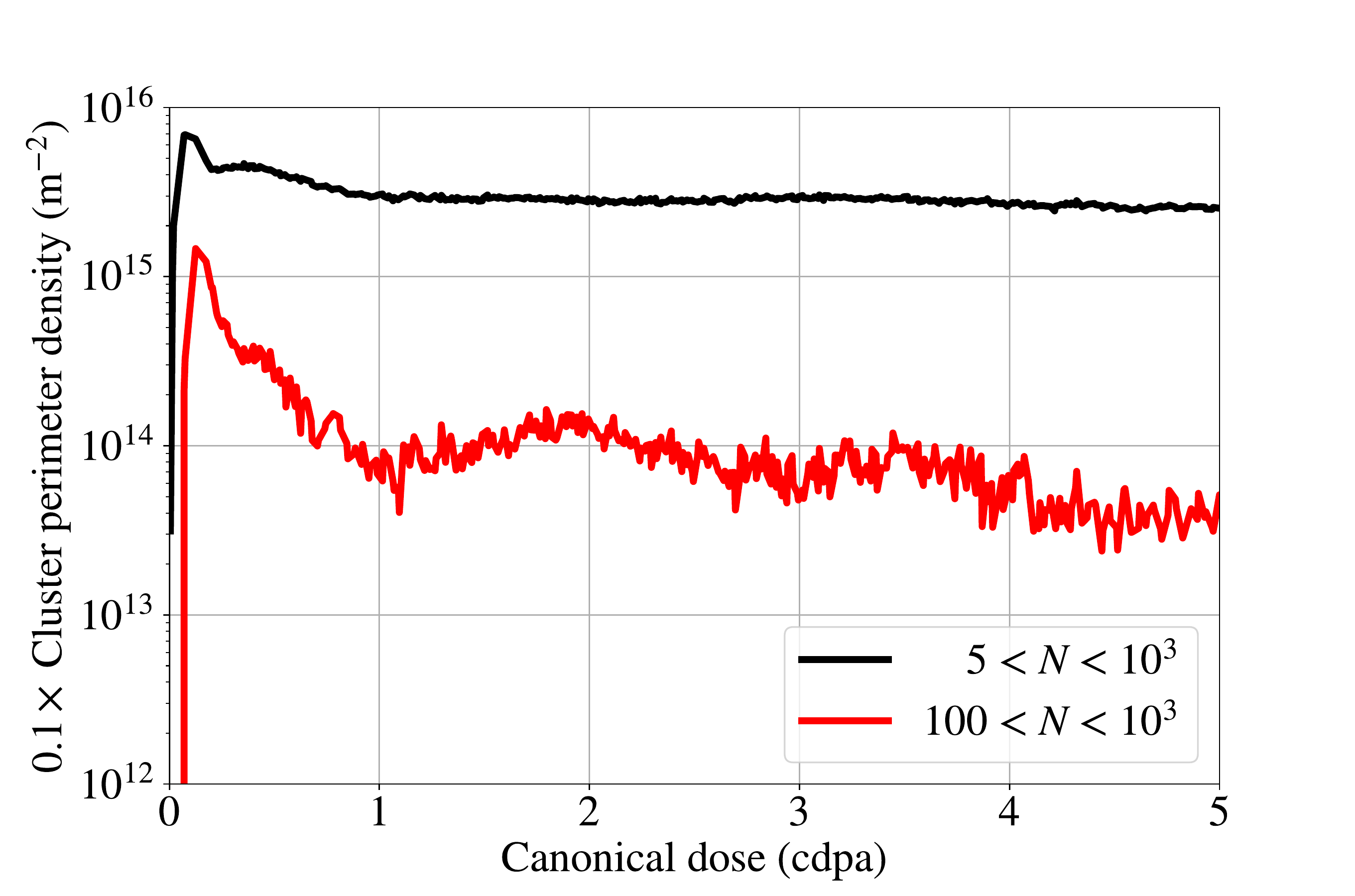}
    \caption{Estimation of dislocation density from interstitial cluster perimeter density. $N$ corresponds to the number of atoms in the cluster and the perimeter of a given cluster is estimated according to Equation \ref{eq:cluster_perim}.}
    \label{fig:SIA_perim_density}
\end{figure}

An immediate observation and the consequence of the power law statistics is that the overwhelming majority of clusters are small. %
Thus, even the order of magnitude of the calculated dislocation density is sensitive to the choice one makes for a minimum detectable size of a dislocation loop.  %
To show this, assume that clusters with cluster population sizes $N$ lying in a range ${ N_{\text{min}} < N < N_{\text{max}}}$ correspond to a circular platelet of interstitial atoms \textit{i.e.}\ a dislocation loop \citep{Gilbert2008}. %
The $N_{\text{min}}$ and $N_{\text{max}}$ thresholds correspond to cluster population sizes that are either sufficiently small to be treated as point defects, or are large enough to be close to the threshold for forming a percolating atomic plane consisting of interconnected dislocation loops. %
Assume that each platelet contains one extra atom per atomic string in the direction of the Burgers vector \citep{Boleininger2018,Boleininger2019}, and involves $N$ atoms with volume $\Omega$ equal to the atomic volume of $\alpha$-Zr. %
Using the formula $({\bf b}\cdot {\bf A})$ for the volume of a dislocation loop, its perimeter can then be estimated as
\begin{equation}
P_S=2\pi \sqrt{{N\Omega \over \pi b}},
\label{eq:cluster_perim}
\end{equation}
where $b$ is the length of the Burgers vector along the normal to the loop habit plane. %
Here, we assume that the platelets form $a$-type dislocation loops such that $b$ is equal to the $a$ lattice parameter. %

Summing the perimeters corresponding to cluster population size distributions between various choices of $N_{\text{min}}$ and $N_{\text{max}}$ provides a measure of the difference in dislocation density due to different choices of thresholds. %
Typically, a dislocation core extends over about five interatomic distances \citep{Boleininger2018, Boleininger2019} and thus one would reasonably argue that $N_{\text{min}}$ should at least be greater than five. %
Furthermore, we should not include the fully formed planes corresponding to a cluster percolating through the periodic boundaries, thus setting $N_{\text{max}} = 10^3$. %
In Figure \ref{fig:SIA_perim_density} we observe an order of magnitude difference between the results involving the counting of all the viable clusters smaller than the population size of the percolating cluster ($5 <N < 10^3$), and the values found by counting all the clusters but increasing the lower threshold to exclude clusters containing less than 100 self-interstitial atoms. %
This example illustrates how the power law distribution of defect sizes in highly irradiated microstructures can profoundly affect how dislocations are counted and their density quantified. %

We may also employ Equation \ref{eq:cluster_perim} to estimate the power law exponent for the distribution of loops with respect to loop diameter $D$ where it is apparent that, assuming that all the loops have the same Burgers vector $b$, $N \propto D^2$. %
Treating $N$ as a continuous variable, we derive the probability density function as a function of loop diameter %
\begin{align}
p_D(D) = \left \lvert \frac{\mathrm{d} N}{\mathrm{d} D} \right \rvert p(N(D)) \propto \frac{1}{D^{2\alpha -1}},
\end{align}
where $\alpha$ is the exponent entering Equation \ref{eq:powlaw}.
Hence the diameter of circular loops is expected to be power law distributed with exponent $\beta = 2\alpha -1$. %
Using the data derived from the CRA simulations in the high dose limit illustrated in Figure \ref{subfig:power_law_exponent_dose} for a 10M atom cell, we find that $\beta = 3.46$. %
In experiments performed by \cite{Ungar2021b} dislocation diameter data measured by TEM were combined with an XRD line profile measurement of the total dislocation density. %
When fitting the dislocation size distribution to a power law the exponent was found to lie in the range $3\le \beta \le 4$, which agrees well with the values derived from our simulations. %

Concluding this section, we note that power law exponent values $\alpha=2$ and $\beta =3$ appear to represent natural low limits characterising the power law statistics of populations of dislocation loops in a heavily irradiated material. Indeed, any power law distribution with $\alpha <2$ would imply a divergent total count of interstitial defects contained in the loops, a paradox that can only be resolved by recognising that loops of very large size are nothing but extra crystallographic atomic planes, which through elastic interactions would tend to modify the finite-size part of the dislocation loop distribution so as to make it normalisable, in this way steering the value of exponent $\alpha$ towards and above the limiting value of 2.

\subsection{Stored energy}
\label{subsec:stored_energy}

Whilst analysing the configuration of atoms helps to describe how the microstructure evolves, on its own this approach provides limited insight into why the observed processes take place. %
Evidently, the repeated creation and relaxation of defects forces the microstructure into a state that has the energy higher than that of a single crystal. %
By examining the excess energy $E_{exc}$ that the system is driven to, we may determine how the accumulation of point defects produces the observed dislocation density. %
For an irradiated microstructure at dose $\phi$ containing $N$ atoms with total energy $E_{total}(\phi)$ we may measure the excess (stored) energy as %
\begin{align}
    E_{exc}(\phi) := E_{total}(\phi) - N E_{coh},
\end{align}
where $E_{coh}$ is the cohesive energy of $\alpha$-Zr. %

Our simulations are carried out under zero global strain boundary conditions. %
Whilst this is computationally convenient, in reality specimens are often irradiated under zero applied stress, allowing the body as a whole to undergo strain. %
Assuming linear elasticity, we may correct for this and remove the stored elastic energy $E_{el}$ from our simulation results. %
The defects produced by radiation damage induce eigenstrains, also known as residual strains, that act as sources of elastic strain. %
Let $\epsilon_{ij}^0$ denote the elastic strain that would come about under zero applied stress boundary conditions. %
In this case, the potential energy of the body is lowered by doing work equal to $\frac{1}{2} \sigma_{ij} \epsilon_{ij}^0$ where Hooke's law determines the stress $\sigma_{ij} = C_{ijkl} \epsilon_{kl}^0$ and the elastic constants tensor is $C_{ijkl}$. %
Enforcing zero global strain requires that the integral of the strain $\epsilon_{ij}$ over the body with volume $V$ is zero or, equivalently, that one has zero volume averaged strain %
\begin{align}
    \langle \epsilon_{ij} \rangle := \frac{1}{V} \int_V \mathrm{d}^3\mathbf{x}'\, \epsilon_{ij}(\mathbf{x}') = 0.
    \label{eq:vol_avg_strain}
\end{align}
This boundary condition is satisfied by the strain %
\begin{align}
    \epsilon_{ij}(\mathbf{x}) = \epsilon_{ij}^0(\mathbf{x}) - \langle \epsilon^0_{ij} \rangle.
    \label{eq:corrected_strain}
\end{align}
Applying Hooke's law to Equation \ref{eq:corrected_strain}, we observe that the system is under a state of global stress $\langle \sigma_{ij} \rangle = C_{ijkl} \langle \epsilon^0_{kl} \rangle$ and we may treat this as the stress that develops in our simulations. %
Therefore, we can compute the stored elastic energy %
\begin{align}
E_{el} = \frac{1}{2} \langle \sigma_{ij} \rangle S_{ijkl} \langle\sigma_{kl} \rangle ,
\end{align}
where $S_{ijkl}$ is the elastic compliance tensor \citep{Warwick2021} related to $C_{ijkl}$ by $S_{ijpq}C_{pqkl} = \frac{1}{2}\left( \delta_{ik}\delta_{jl} + \delta_{il}\delta_{jk}\right)$. %
We find across all the employed potentials that $E_{el}$ accounts only for a small fraction $< 10 \%$ of the stored energy, meaning that the vast majority of stored energy is contained in the point defect centres, dislocation cores and fluctuations arising from the elastic fields of these defects. %
Indeed, when a high dose snapshot was explicitly relaxed under zero pressure we found that the difference in excess energy to that corrected for by our estimate of $E_{el}$ is of the same order of magnitude. %
Furthermore, the microstructure did not change significantly providing more indication of the minor role of the global elastic energy. %

\begin{figure}
    \begin{subfigure}{.45\textwidth}
        \centering
        \includegraphics[width=.8\linewidth]{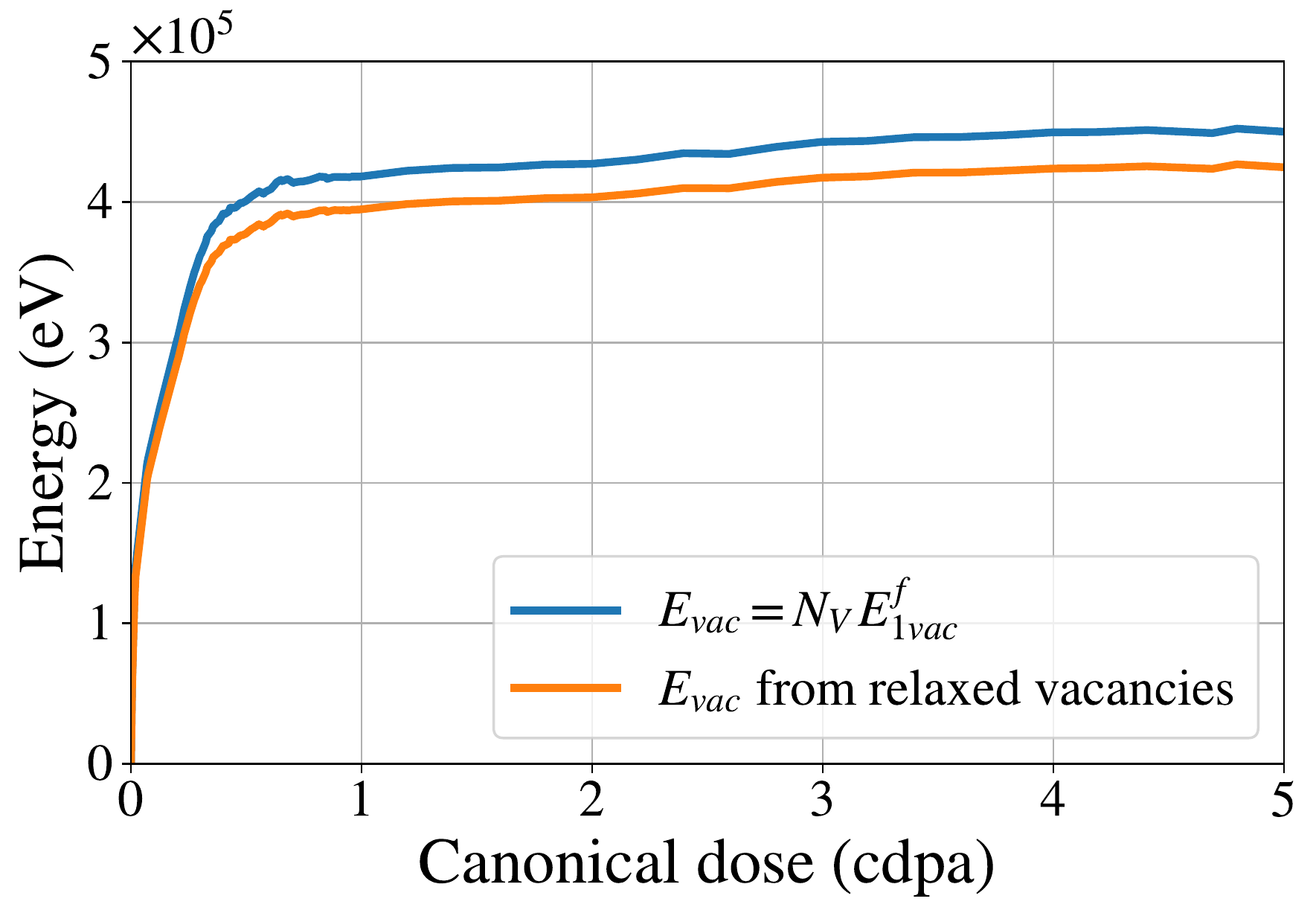}
        \caption{}
        \label{subfig:vac_comparison}
    \end{subfigure}
    \begin{subfigure}{.45\textwidth}
        \centering
        \includegraphics[width=.8\linewidth]{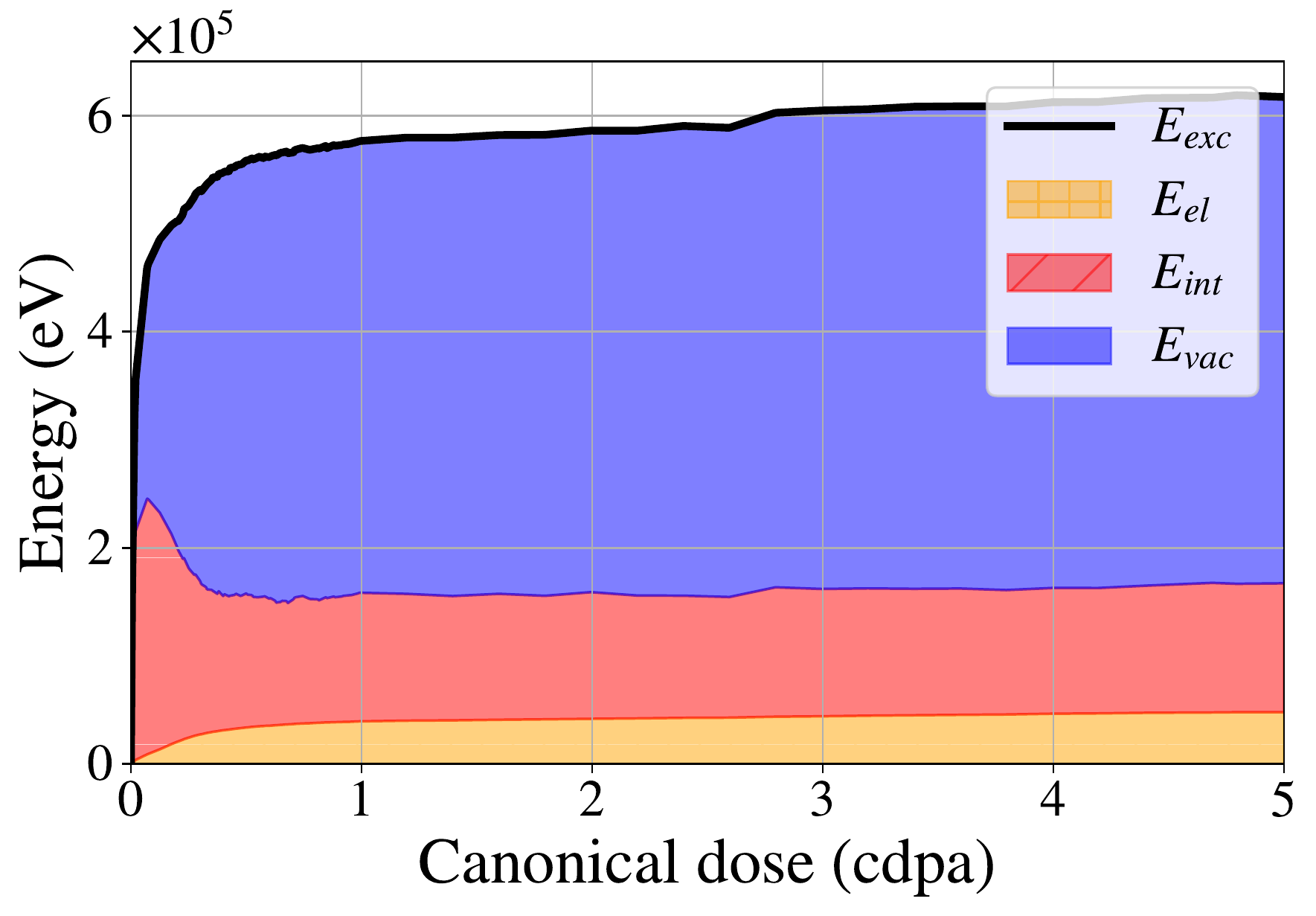}
        \caption{}
        \label{subfig:energy_breakdown}
    \end{subfigure} \\
    \begin{subfigure}{\textwidth}
        \centering
        \includegraphics[width=.7\linewidth]{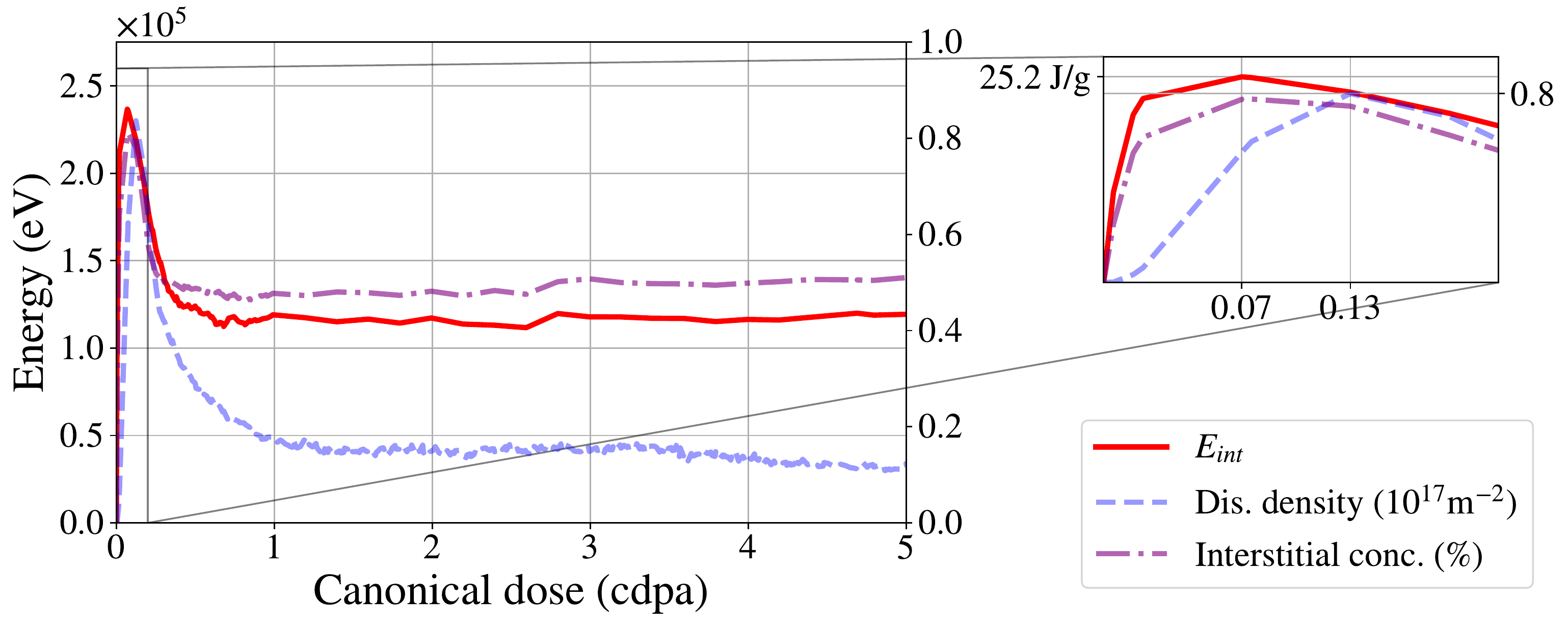}
        \caption{}
        \label{subfig:E_int_corr}
    \end{subfigure}
    \caption{Contributions of microstructural defects to the excess energy. Results shown are for the MA1 potential using 10 million atoms. \ref{subfig:vac_comparison}: Comparison of vacancy energy $E_{vac}$ estimated by assuming all $N_V$ vacancies are isolated with formation energy $E^f_{1vac}$ (blue curve) with $E_{vac}$ calculated \textit{via} explicitly relaxing pristine $\alpha$-Zr containing the same number and arrangement of vacancies. \ref{subfig:energy_breakdown} : Breakdown of total excess energy $E_{exc}$ into elastic $E_{el}$, interstitial $E_{int}$ and vacancy $E_{vac}$ contributions. \ref{subfig:E_int_corr}: Comparison of interstitial excess energy, concentration and dislocation density that exhibits the similarity in profile between all three quantities. A peak occurs in all three profiles near \SI{0.1}{cdpa}. }
    \label{figs:energies}
\end{figure}

The remaining stored energy is confined in the two classes of defects associated with vacancies and interstitials. %
Given that vacancies do not cluster significantly, we may estimate their energetic contribution as %
\begin{align}
	E_{vac}(\phi) = N_{vac}(\phi) E^f_{1vac},
 \label{eq:vac_en}
\end{align}where, for dose $\phi$, the number of vacancies is denoted $N_{vac}$ and $E^{f}_{1vac}$ is the formation energy of a single vacancy. %
Values of $E^{f}_{1vac}$ for each of the potentials used in this study may be found in \cite{Mendelev2007}. %
To check the validity of Equation \ref{eq:vac_en} we isolated the $N_{vac}$ vacancies identified by Wigner-Seitz analysis and subsequently relaxed an $N$-atom supercell of pristine $\alpha$-Zr containing the same number of vacancies in the same positions. %
The simulation cell was relaxed and the formation energy was computed as %
\begin{align}
	E_{vac} = E^{total}_{vac} - (N - N_{vac}) E_{coh},
\end{align}
where $E^{total}_{vac}$ is the resulting total energy. %
Figure \ref{subfig:vac_comparison} summarises this process and shows the resulting formation energy for these arrangements of vacancies. %
Thus we observe that Equation \ref{eq:vac_en} is a good approximation, providing further evidence that the majority of $E_{vac}$ is due to isolated vacancies. %

The contribution to $E_{exc}$ from the formation energy of small interstitial clusters and dislocation content may now be calculated as %
\begin{align}
	E_{int} = E_{exc} - E_{el} - E_{vac}.
\end{align}
In Figure \ref{subfig:energy_breakdown} we show the relative proportion of elastic, interstitial and vacancy contributions to the total excess energy where it is evident that the elastic contribution is in the minority, vacancy contributions dominate and the interstitial contribution follows a similar trend to the dislocation density profile shown in Figure \ref{fig:CRA_XRD_comparison}. %
The profile of $E_{int}$ is shown in Figure \ref{subfig:E_int_corr} together with the interstitial concentration and dislocation density in order to highlight their correlation with each other. %

Recently, Differential Scanning Calorimetry (DSC)  experiments were performed to measure the stored energy of irradiated titanium as a means of inferring the number of defects present \citep{Hirst2022}. %
The measurements indicated stored energies associated with irradiation induced defects to be on the order of \SI{0.1}{J g^{-1}}. %
Their analysis also allowed the authors to infer the presence of defects that are invisible to TEM imaging. %
At \SI{0.07}{cdpa} in Figure \ref{subfig:E_int_corr}, we find that the peak in specific energy associated with $E_{int}$ is \SI{25}{J g^{-1}} that subsequently drops and plateaus at high dose to \SI{12}{J g^{-1}}. %
At elevated temperatures, the defect content is likely to be $\sim 10 \%$ of that calculated in our simulations cf. \citep{Mason2020,Mason2021} and thus we expect the associated stored energy in Zr to be comparable to \SI{1}{J g^{-1}}. %

\section{Conclusions}
\label{sec:conclusions}

In summary, we have performed experiments and simulations showing that the dislocation density in irradiated zirconium and zircaloys exhibits a peak at a moderate dose and then saturates at doses greater than \SI{1}{dpa}. %
Simulations indicate that this occurs in a regime of dose rate and temperature where microstructural evolution is predominantly driven by stress relaxation. %
The material enters a critical state at \SI{\sim 1}{dpa}, where interstitial clusters grow to a sufficient size to percolate the volume of the material. %
At high dose, the population of smaller clusters and dislocation loops is distributed as a function of cluster defect content according to a power law statistics with the exponent close to $\alpha \approx 2.2$. As a function of defect diameter, this results in a power law distribution of defect clusters with exponent of $\beta \approx 3.5$, which compares favourably with the range of values $3\le \beta \le 4$ derived from experimental observations \citep{Ungar2021b}.%
The analysis highlights the significance of precise definition of defect sizes included in the measured dislocation densities. Irrespectively of the statistics of dislocation structures, the trend in the dislocation density evolution in zirconium irradiated at temperatures below $\sim 350^\circ \: \mathrm{C}$ is clear, the dislocation density saturates as a function of dose. %

\section{Acknowledgements}

This work received funding from the RCUK
Energy Programme Grant No. EP/W006839/1 and MIDAS EPSRC Grant No. EP/S01702X/1, and was partially carried out within the framework of the EUROfusion Consortium,funded by the European Union via the Euratom Research and Training Programme (Grant Agreement No 101052200 — EUROfusion). Views and opinions expressed are however those of the authors only and do not necessarily reflect those of the European Union or the European Commission. Neither the European Union nor the European Commission can be held responsible for them. We acknowledge DESY (Hamburg, Germany), a member of the Helmholtz Association HGF, for the provision of experimental facilities. We gratefully acknowledge the use of the high-performance computing facility MARCONI (Bologna, Italy) provided by EUROfusion, and computing resources supplied by the IRIS (STFC) Consortium. This work also received support from the EPSRC Access to HPC Programme on the ARCHER2 UK National Supercomputing Service (http://www.archer2.ac.uk).

\printcredits

\bibliographystyle{cas-model2-names}

\bibliography{references}

\begin{thebibliography}{84}
\expandafter\ifx\csname natexlab\endcsname\relax\def\natexlab#1{#1}\fi
\providecommand{\url}[1]{\texttt{#1}}
\providecommand{\href}[2]{#2}
\providecommand{\path}[1]{#1}
\providecommand{\DOIprefix}{doi:}
\providecommand{\ArXivprefix}{arXiv:}
\providecommand{\URLprefix}{URL: }
\providecommand{\Pubmedprefix}{pmid:}
\providecommand{\doi}[1]{\href{http://dx.doi.org/#1}{\path{#1}}}
\providecommand{\Pubmed}[1]{\href{pmid:#1}{\path{#1}}}
\providecommand{\bibinfo}[2]{#2}
\ifx\xfnm\relax \def\xfnm[#1]{\unskip,\space#1}\fi
\bibitem[{Adamson et~al.(2019)Adamson, Coleman and Griffiths}]{Adamson2019}
\bibinfo{author}{Adamson, R.B.}, \bibinfo{author}{Coleman, C.E.},
  \bibinfo{author}{Griffiths, M.}, \bibinfo{year}{2019}.
\newblock \bibinfo{title}{{Irradiation creep and growth of zirconium alloys: A
  critical review}}.
\newblock \bibinfo{journal}{Journal of Nuclear Materials}
  \bibinfo{volume}{521}, \bibinfo{pages}{167--244}.
\newblock \DOIprefix\doi{10.1016/j.jnucmat.2019.04.021}.
\bibitem[{Ahrens et~al.(2005)Ahrens, Geveci and Law}]{AHRENS2005}
\bibinfo{author}{Ahrens, J.}, \bibinfo{author}{Geveci, B.},
  \bibinfo{author}{Law, C.}, \bibinfo{year}{2005}.
\newblock \bibinfo{title}{{ParaView: An End-User Tool for Large-Data
  Visualization}}, in: \bibinfo{editor}{Hansen, C.D.},
  \bibinfo{editor}{Johnson, C.R.} (Eds.), \bibinfo{booktitle}{Visualization
  Handbook}. \bibinfo{publisher}{Elsevier}, pp. \bibinfo{pages}{717--731}.
\newblock \DOIprefix\doi{10.1016/B978-012387582-2/50038-1}.
\bibitem[{Allnatt and Lidiard(1993)}]{Allnatt1993}
\bibinfo{author}{Allnatt, A.R.}, \bibinfo{author}{Lidiard, A.B.},
  \bibinfo{year}{1993}.
\newblock \bibinfo{title}{Atomic Transport in Solids}.
\newblock \bibinfo{publisher}{Cambridge University Press, Cambridge, England}.
\newblock \DOIprefix\doi{10.1017/CBO9780511563904}.
\bibitem[{Alstott et~al.(2014)Alstott, Bullmore and Plenz}]{Alstott2014}
\bibinfo{author}{Alstott, J.}, \bibinfo{author}{Bullmore, E.},
  \bibinfo{author}{Plenz, D.}, \bibinfo{year}{2014}.
\newblock \bibinfo{title}{{powerlaw: A Python Package for Analysis of
  Heavy-Tailed Distributions}}.
\newblock \bibinfo{journal}{PLoS ONE} \bibinfo{volume}{9},
  \bibinfo{pages}{e85777}.
\newblock \DOIprefix\doi{10.1371/journal.pone.0085777}.
\bibitem[{Arakawa et~al.(2020)Arakawa, Marinica, Fitzgerald, Proville,
  Nguyen-Manh, Dudarev, Ma, Swinburne, Goryaeva, Yamada, Amino, Arai, Yamamoto,
  Higuchi, Tanaka, Yasuda, Yasuda and Mori}]{Arakawa2020}
\bibinfo{author}{Arakawa, K.}, \bibinfo{author}{Marinica, M.C.},
  \bibinfo{author}{Fitzgerald, S.}, \bibinfo{author}{Proville, L.},
  \bibinfo{author}{Nguyen-Manh, D.}, \bibinfo{author}{Dudarev, S.L.},
  \bibinfo{author}{Ma, P.W.}, \bibinfo{author}{Swinburne, T.D.},
  \bibinfo{author}{Goryaeva, A.M.}, \bibinfo{author}{Yamada, T.},
  \bibinfo{author}{Amino, T.}, \bibinfo{author}{Arai, S.},
  \bibinfo{author}{Yamamoto, Y.}, \bibinfo{author}{Higuchi, K.},
  \bibinfo{author}{Tanaka, N.}, \bibinfo{author}{Yasuda, H.},
  \bibinfo{author}{Yasuda, T.}, \bibinfo{author}{Mori, H.},
  \bibinfo{year}{2020}.
\newblock \bibinfo{title}{Quantum de-trapping and transport of heavy defects in
  tungsten}.
\newblock \bibinfo{journal}{Nature Materials} \bibinfo{volume}{19},
  \bibinfo{pages}{508}.
\newblock \DOIprefix\doi{10.1038/s41563-019-0584-0}.
\bibitem[{Arsenlis and Parks(1999)}]{Arsenlis1999}
\bibinfo{author}{Arsenlis, A.}, \bibinfo{author}{Parks, D.},
  \bibinfo{year}{1999}.
\newblock \bibinfo{title}{{Crystallographic aspects of geometrically-necessary
  and statistically-stored dislocation density}}.
\newblock \bibinfo{journal}{Acta Materialia} \bibinfo{volume}{47},
  \bibinfo{pages}{1597--1611}.
\newblock \DOIprefix\doi{10.1016/S1359-6454(99)00020-8}.
\bibitem[{Balogh et~al.(2016)Balogh, Long and Daymond}]{Balogh2016}
\bibinfo{author}{Balogh, L.}, \bibinfo{author}{Long, F.},
  \bibinfo{author}{Daymond, M.R.}, \bibinfo{year}{2016}.
\newblock \bibinfo{title}{Contrast factors of irradiation-induced dislocation
  loops in hexagonal materials}.
\newblock \bibinfo{journal}{Journal of Applied Crystallography}
  \bibinfo{volume}{49}, \bibinfo{pages}{2184--2200}.
\newblock \DOIprefix\doi{10.1107/S1600576716018136}.
\bibitem[{Bitzek et~al.(2006)Bitzek, Koskinen, G{\"{a}}hler, Moseler and
  Gumbsch}]{Bitzek2006}
\bibinfo{author}{Bitzek, E.}, \bibinfo{author}{Koskinen, P.},
  \bibinfo{author}{G{\"{a}}hler, F.}, \bibinfo{author}{Moseler, M.},
  \bibinfo{author}{Gumbsch, P.}, \bibinfo{year}{2006}.
\newblock \bibinfo{title}{{Structural Relaxation Made Simple}}.
\newblock \bibinfo{journal}{Physical Review Letters} \bibinfo{volume}{97},
  \bibinfo{pages}{170201}.
\newblock \DOIprefix\doi{10.1103/PhysRevLett.97.170201}.
\bibitem[{Boleininger and Dudarev(2019)}]{Boleininger2019}
\bibinfo{author}{Boleininger, M.}, \bibinfo{author}{Dudarev, S.L.},
  \bibinfo{year}{2019}.
\newblock \bibinfo{title}{Continuum model for the core of a straight mixed
  dislocation}.
\newblock \bibinfo{journal}{Physical Review Materials} \bibinfo{volume}{3},
  \bibinfo{pages}{093801}.
\newblock \DOIprefix\doi{10.1103/PhysRevMaterials.3.093801}.
\bibitem[{Boleininger et~al.(2022)Boleininger, Dudarev, Mason and
  Martínez}]{Boleininger2022}
\bibinfo{author}{Boleininger, M.}, \bibinfo{author}{Dudarev, S.L.},
  \bibinfo{author}{Mason, D.R.}, \bibinfo{author}{Martínez, E.},
  \bibinfo{year}{2022}.
\newblock \bibinfo{title}{Volume of a dislocation network}.
\newblock \bibinfo{journal}{Physical Review Materials} \bibinfo{volume}{6},
  \bibinfo{pages}{063601}.
\newblock \DOIprefix\doi{10.1103/PhysRevMaterials.6.063601}.
\bibitem[{Boleininger et~al.(2023)Boleininger, Mason, Sand and
  Dudarev}]{Boleininger2023}
\bibinfo{author}{Boleininger, M.}, \bibinfo{author}{Mason, D.R.},
  \bibinfo{author}{Sand, A.E.}, \bibinfo{author}{Dudarev, S.L.},
  \bibinfo{year}{2023}.
\newblock \bibinfo{title}{Microstructure of a heavily irradiated metal exposed
  to a spectrum of atomic recoils}.
\newblock \bibinfo{journal}{Scientific Reports} \bibinfo{volume}{13},
  \bibinfo{pages}{1684}.
\newblock \DOIprefix\doi{10.1038/s41598-022-27087-w}.
\bibitem[{Boleininger et~al.(2018)Boleininger, Swinburne and
  Dudarev}]{Boleininger2018}
\bibinfo{author}{Boleininger, M.}, \bibinfo{author}{Swinburne, T.D.},
  \bibinfo{author}{Dudarev, S.L.}, \bibinfo{year}{2018}.
\newblock \bibinfo{title}{{Atomistic-to-continuum description of edge
  dislocation core: Unification of the Peierls-Nabarro model with linear
  elasticity}}.
\newblock \bibinfo{journal}{Physical Review Materials} \bibinfo{volume}{2},
  \bibinfo{pages}{083803}.
\newblock \DOIprefix\doi{10.1103/PhysRevMaterials.2.083803}.
\bibitem[{Chartier and Marinica(2019)}]{Chartier2019}
\bibinfo{author}{Chartier, A.}, \bibinfo{author}{Marinica, M.C.},
  \bibinfo{year}{2019}.
\newblock \bibinfo{title}{Rearrangement of interstitial defects in alpha-{Fe}
  under extreme condition}.
\newblock \bibinfo{journal}{Acta Materialia} \bibinfo{volume}{180},
  \bibinfo{pages}{141--148}.
\newblock \DOIprefix\doi{10.1016/j.actamat.2019.09.007}.
\bibitem[{Choi and Kim(2013)}]{CHOI2013}
\bibinfo{author}{Choi, S.I.}, \bibinfo{author}{Kim, J.H.},
  \bibinfo{year}{2013}.
\newblock \bibinfo{title}{Radiation-induced dislocation and growth behavior of
  zirconium and zirconium alloys - a review}.
\newblock \bibinfo{journal}{Nuclear Engineering and Technology}
  \bibinfo{volume}{45}, \bibinfo{pages}{385--392}.
\newblock \DOIprefix\doi{10.5516/NET.07.2013.035}.
\bibitem[{Christensen et~al.(2020)Christensen, Wolf, Freeman, Wimmer, Adamson,
  Griffiths and Mader}]{Christensen2020}
\bibinfo{author}{Christensen, M.}, \bibinfo{author}{Wolf, W.},
  \bibinfo{author}{Freeman, C.}, \bibinfo{author}{Wimmer, E.},
  \bibinfo{author}{Adamson, R.}, \bibinfo{author}{Griffiths, M.},
  \bibinfo{author}{Mader, E.}, \bibinfo{year}{2020}.
\newblock \bibinfo{title}{{Vacancy loops in Breakaway Irradiation Growth of
  zirconium: Insight from atomistic simulations}}.
\newblock \bibinfo{journal}{Journal of Nuclear Materials}
  \bibinfo{volume}{529}, \bibinfo{pages}{151946}.
\newblock \DOIprefix\doi{10.1016/j.jnucmat.2019.151946}.
\bibitem[{Das et~al.(2018)Das, Hofmann and Tarleton}]{Das2018}
\bibinfo{author}{Das, S.}, \bibinfo{author}{Hofmann, F.},
  \bibinfo{author}{Tarleton, E.}, \bibinfo{year}{2018}.
\newblock \bibinfo{title}{{Consistent determination of geometrically necessary
  dislocation density from simulations and experiments}}.
\newblock \bibinfo{journal}{International Journal of Plasticity}
  \bibinfo{volume}{109}, \bibinfo{pages}{18--42}.
\newblock \DOIprefix\doi{10.1016/j.ijplas.2018.05.001}.
\bibitem[{Debye(1915)}]{Debye1915}
\bibinfo{author}{Debye, P.}, \bibinfo{year}{1915}.
\newblock \bibinfo{title}{{Zerstreuung von R{\"{o}}ntgenstrahlen}}.
\newblock \bibinfo{journal}{Annalen der Physik} \bibinfo{volume}{351},
  \bibinfo{pages}{809--823}.
\newblock \DOIprefix\doi{10.1002/andp.19153510606}.
\bibitem[{Derlet and Dudarev(2020)}]{DerletDudarev2020}
\bibinfo{author}{Derlet, P.M.}, \bibinfo{author}{Dudarev, S.L.},
  \bibinfo{year}{2020}.
\newblock \bibinfo{title}{{Microscopic structure of a heavily irradiated
  material}}.
\newblock \bibinfo{journal}{Physical Review Materials} \bibinfo{volume}{4},
  \bibinfo{pages}{023605}.
\newblock \DOIprefix\doi{10.1103/PhysRevMaterials.4.023605}.
\bibitem[{Domain and Legris(2005)}]{Domain2005}
\bibinfo{author}{Domain, C.}, \bibinfo{author}{Legris, A.},
  \bibinfo{year}{2005}.
\newblock \bibinfo{title}{Ab initio atomic-scale determination of point-defect
  structure in hcp zirconium}.
\newblock \bibinfo{journal}{Philosophical Magazine} \bibinfo{volume}{85},
  \bibinfo{pages}{569--575}.
\newblock \DOIprefix\doi{10.1080/14786430412331334625}.
\bibitem[{Dudarev(2008)}]{Dudarev2008}
\bibinfo{author}{Dudarev, S.L.}, \bibinfo{year}{2008}.
\newblock \bibinfo{title}{The non-{Arrhenius} migration of interstitial defects
  in bcc transition metals}.
\newblock \bibinfo{journal}{Comptes Rendus Physique} \bibinfo{volume}{9},
  \bibinfo{pages}{409 -- 417}.
\newblock \DOIprefix\doi{10.1016/j.crhy.2007.09.019}.
\bibitem[{Dudarev(2013)}]{Dudarev2013}
\bibinfo{author}{Dudarev, S.L.}, \bibinfo{year}{2013}.
\newblock \bibinfo{title}{Density functional theory models for radiation
  damage}.
\newblock \bibinfo{journal}{Annu. Rev. Mater. Res.} \bibinfo{volume}{43},
  \bibinfo{pages}{35 -- 61}.
\newblock \DOIprefix\doi{10.1146/annurev-matsci-071312-121626}.
\bibitem[{Dudarev et~al.(2010)Dudarev, Gilbert, Arakawa, Mori, Yao, Jenkins and
  Derlet}]{Dudarev2010}
\bibinfo{author}{Dudarev, S.L.}, \bibinfo{author}{Gilbert, M.R.},
  \bibinfo{author}{Arakawa, K.}, \bibinfo{author}{Mori, H.},
  \bibinfo{author}{Yao, Z.}, \bibinfo{author}{Jenkins, M.L.},
  \bibinfo{author}{Derlet, P.M.}, \bibinfo{year}{2010}.
\newblock \bibinfo{title}{{Langevin} model for real-time {Brownian} dynamics of
  interacting nanodefects in irradiated metals}.
\newblock \bibinfo{journal}{Physical Review B} \bibinfo{volume}{81},
  \bibinfo{pages}{224107}.
\newblock \DOIprefix\doi{10.1103/PhysRevB.81.224107}.
\bibitem[{Dudarev and Ma(2018)}]{Dudarev2018}
\bibinfo{author}{Dudarev, S.L.}, \bibinfo{author}{Ma, P.W.},
  \bibinfo{year}{2018}.
\newblock \bibinfo{title}{{Elastic fields, dipole tensors, and interaction
  between self-interstitial atom defects in bcc transition metals}}.
\newblock \bibinfo{journal}{Physical Review Materials} \bibinfo{volume}{2},
  \bibinfo{pages}{033602}.
\newblock \DOIprefix\doi{10.1103/PhysRevMaterials.2.033602}.
\bibitem[{Ehrhart et~al.(1991)Ehrhart, Jung, Schultz and
  Ullmaier}]{LandoltBornstein1991}
\bibinfo{author}{Ehrhart, P.}, \bibinfo{author}{Jung, P.},
  \bibinfo{author}{Schultz, H.}, \bibinfo{author}{Ullmaier, H.},
  \bibinfo{year}{1991}.
\newblock \bibinfo{title}{Landolt-B{\"o}rnstein - Group III Condensed Matter
  {\textperiodcentered} Volume 25: ``Atomic Defects in Metals''}.
\newblock \bibinfo{publisher}{Springer-Verlag Berlin Heidelberg}.
\newblock \DOIprefix\doi{10.1007/10011948_45}.
\bibitem[{Faken and J{\'{o}}nsson(1994)}]{Faken1994}
\bibinfo{author}{Faken, D.}, \bibinfo{author}{J{\'{o}}nsson, H.},
  \bibinfo{year}{1994}.
\newblock \bibinfo{title}{{Systematic analysis of local atomic structure
  combined with 3D computer graphics}}.
\newblock \bibinfo{journal}{Computational Materials Science}
  \bibinfo{volume}{2}, \bibinfo{pages}{279--286}.
\newblock \DOIprefix\doi{10.1016/0927-0256(94)90109-0}.
\bibitem[{Fu et~al.(2005)Fu, {Dalla Torre}, Willaime, Bocquet and
  Barbu}]{Fu2005}
\bibinfo{author}{Fu, C.C.}, \bibinfo{author}{{Dalla Torre}, J.},
  \bibinfo{author}{Willaime, F.}, \bibinfo{author}{Bocquet, J.L.},
  \bibinfo{author}{Barbu, A.}, \bibinfo{year}{2005}.
\newblock \bibinfo{title}{Multiscale modelling of defect kinetics in irradiated
  iron}.
\newblock \bibinfo{journal}{Nature Materials} \bibinfo{volume}{4},
  \bibinfo{pages}{68 -- 74}.
\newblock \DOIprefix\doi{10.1038/nmat1286}.
\bibitem[{Fu et~al.(2008)Fu, Meslin, Barbu, Willaime and Oison}]{Fu2008}
\bibinfo{author}{Fu, C.C.}, \bibinfo{author}{Meslin, E.},
  \bibinfo{author}{Barbu, A.}, \bibinfo{author}{Willaime, F.},
  \bibinfo{author}{Oison, V.}, \bibinfo{year}{2008}.
\newblock \bibinfo{title}{Effect of {C} on vacancy migration in $\alpha$-iron}.
\newblock \bibinfo{journal}{Solid State Phenomena} \bibinfo{volume}{139},
  \bibinfo{pages}{157 -- 164}.
\newblock \DOIprefix\doi{10.4028/www.scientific.net/SSP.139.157}.
\bibitem[{Gilbert et~al.(2008)Gilbert, Dudarev, Derlet and
  Pettifor}]{Gilbert2008}
\bibinfo{author}{Gilbert, M.R.}, \bibinfo{author}{Dudarev, S.L.},
  \bibinfo{author}{Derlet, P.M.}, \bibinfo{author}{Pettifor, D.G.},
  \bibinfo{year}{2008}.
\newblock \bibinfo{title}{Structure and metastability of mesoscopic vacancy and
  interstitial loop defects in iron and tungsten}.
\newblock \bibinfo{journal}{J. Phys.: Condens. Matter} \bibinfo{volume}{20},
  \bibinfo{pages}{345214}.
\newblock \DOIprefix\doi{10.1088/0953-8984/20/34/345214}.
\bibitem[{Granberg et~al.(2023)Granberg, Mason and Byggmästar}]{Granberg2023}
\bibinfo{author}{Granberg, F.}, \bibinfo{author}{Mason, D.},
  \bibinfo{author}{Byggmästar, J.}, \bibinfo{year}{2023}.
\newblock \bibinfo{title}{Effect of simulation technique on the high-dose
  damage in tungsten}.
\newblock \bibinfo{journal}{Computational Materials Science}
  \bibinfo{volume}{217}, \bibinfo{pages}{111902}.
\newblock \DOIprefix\doi{10.1016/j.commatsci.2022.111902}.
\bibitem[{Griffiths(2020)}]{Griffiths2020}
\bibinfo{author}{Griffiths, M.}, \bibinfo{year}{2020}.
\newblock \bibinfo{title}{{1.11 - Irradiation Growth}}, in:
  \bibinfo{editor}{Konings, R.J.}, \bibinfo{editor}{Stoller, R.E.} (Eds.),
  \bibinfo{booktitle}{Comprehensive Nuclear Materials}.
  \bibinfo{edition}{second} ed.. \bibinfo{publisher}{Elsevier},
  \bibinfo{address}{Oxford}. volume~\bibinfo{volume}{1}, pp.
  \bibinfo{pages}{367--405}.
\newblock \DOIprefix\doi{10.1016/B978-0-12-803581-8.11646-7}.
\bibitem[{Groma and Borbély(2004)}]{Groma2004}
\bibinfo{author}{Groma, I.}, \bibinfo{author}{Borbély, A.},
  \bibinfo{year}{2004}.
\newblock \bibinfo{title}{X-ray {Peak} {Broadening} {Due} to {Inhomogeneous}
  {Dislocation} {Distributions}}, in: \bibinfo{editor}{Mittemeijer, E.J.},
  \bibinfo{editor}{Scardi, P.} (Eds.), \bibinfo{booktitle}{Diffraction
  {Analysis} of the {Microstructure} of {Materials}}.
  \bibinfo{publisher}{Springer}, \bibinfo{address}{Berlin, Heidelberg}.
  Springer {Series} in {Materials} {Science}, pp. \bibinfo{pages}{287--307}.
\newblock \DOIprefix\doi{10.1007/978-3-662-06723-9_11}.
\bibitem[{Heynsworth and Goldberg(1965)}]{Abramovitz}
\bibinfo{author}{Heynsworth, E.Y.}, \bibinfo{author}{Goldberg, K.},
  \bibinfo{year}{1965}.
\newblock \bibinfo{title}{{Bernoulli and Euler Polynomials, Riemann Zeta
  Function}}, in: \bibinfo{editor}{Abramovitz, M.}, \bibinfo{editor}{Stegun,
  I.} (Eds.), \bibinfo{booktitle}{Handbook of Mathematical Functions}.
  \bibinfo{publisher}{Dover, New York}, pp. \bibinfo{pages}{803 -- 819}.
\bibitem[{Hirst et~al.(2022)Hirst, Granberg, Kombaiah, Cao, Middlemas, Kemp,
  Li, Nordlund and Short}]{Hirst2022}
\bibinfo{author}{Hirst, C.A.}, \bibinfo{author}{Granberg, F.},
  \bibinfo{author}{Kombaiah, B.}, \bibinfo{author}{Cao, P.},
  \bibinfo{author}{Middlemas, S.}, \bibinfo{author}{Kemp, R.S.},
  \bibinfo{author}{Li, J.}, \bibinfo{author}{Nordlund, K.},
  \bibinfo{author}{Short, M.P.}, \bibinfo{year}{2022}.
\newblock \bibinfo{title}{{Revealing hidden defects through stored energy
  measurements of radiation damage}}.
\newblock \bibinfo{journal}{Science Advances} \bibinfo{volume}{8}.
\newblock \DOIprefix\doi{10.1126/sciadv.abn2733}.
\bibitem[{Holt(1988)}]{Holt1988}
\bibinfo{author}{Holt, R.}, \bibinfo{year}{1988}.
\newblock \bibinfo{title}{{Mechanisms of irradiation growth of alpha-zirconium
  alloys}}.
\newblock \bibinfo{journal}{Journal of Nuclear Materials}
  \bibinfo{volume}{159}, \bibinfo{pages}{310--338}.
\newblock \DOIprefix\doi{10.1016/0022-3115(88)90099-2}.
\bibitem[{Hull and Bacon(2011)}]{Hull2011}
\bibinfo{author}{Hull, D.}, \bibinfo{author}{Bacon, D.}, \bibinfo{year}{2011}.
\newblock \bibinfo{title}{Introduction to Dislocations : Chapter 1 - Defects in
  Crystals}.
\newblock \bibinfo{edition}{fifth} ed.,
  \bibinfo{publisher}{Butterworth-Heinemann}, \bibinfo{address}{Oxford}.
\newblock \DOIprefix\doi{10.1016/B978-0-08-096672-4.00001-3}.
\bibitem[{Jones et~al.(2016)Jones, Zimmerman and Po}]{Jones2016}
\bibinfo{author}{Jones, R.E.}, \bibinfo{author}{Zimmerman, J.A.},
  \bibinfo{author}{Po, G.}, \bibinfo{year}{2016}.
\newblock \bibinfo{title}{{Comparison of Dislocation Density Tensor Fields
  Derived from Discrete Dislocation Dynamics and Crystal Plasticity Simulations
  of Torsion}}.
\newblock \bibinfo{journal}{Journal of Materials Science Research}
  \bibinfo{volume}{5}, \bibinfo{pages}{44}.
\newblock \DOIprefix\doi{10.5539/jmsr.v5n4p44}.
\bibitem[{Kabir et~al.(2010)Kabir, Lau, Lin, Yip and {Van Vliet}}]{Kabir2010}
\bibinfo{author}{Kabir, M.}, \bibinfo{author}{Lau, T.T.}, \bibinfo{author}{Lin,
  X.}, \bibinfo{author}{Yip, S.}, \bibinfo{author}{{Van Vliet}, K.J.},
  \bibinfo{year}{2010}.
\newblock \bibinfo{title}{Effects of vacancy-solute clusters on diffusivity in
  metastable {Fe-C} alloys}.
\newblock \bibinfo{journal}{Physical Review B} \bibinfo{volume}{82},
  \bibinfo{pages}{134112}.
\newblock \DOIprefix\doi{10.1103/PhysRevB.82.134112}.
\bibitem[{Kamminga and Delhez(2000)}]{Kamminga2000}
\bibinfo{author}{Kamminga, J.D.}, \bibinfo{author}{Delhez, R.},
  \bibinfo{year}{2000}.
\newblock \bibinfo{title}{Calculation of diffraction line profiles from
  specimens with dislocations. {A} comparison of analytical models with
  computer simulations}.
\newblock \bibinfo{journal}{Journal of Applied Crystallography}
  \bibinfo{volume}{33}, \bibinfo{pages}{1122--1127}.
\newblock \DOIprefix\doi{10.1107/S0021889800006750}.
\bibitem[{Landauer and Swanson(1961)}]{Landauer1961}
\bibinfo{author}{Landauer, R.}, \bibinfo{author}{Swanson, J.A.},
  \bibinfo{year}{1961}.
\newblock \bibinfo{title}{Frequency factors in the thermally activated
  processes}.
\newblock \bibinfo{journal}{Physical Review} \bibinfo{volume}{121},
  \bibinfo{pages}{1668 -- 1674}.
\newblock \DOIprefix\doi{10.1103/PhysRev.121.1668}.
\bibitem[{Lemaignan(2012)}]{Lemaignan2012}
\bibinfo{author}{Lemaignan, C.}, \bibinfo{year}{2012}.
\newblock \bibinfo{title}{{2.07 - Zirconium Alloys: Properties and
  Characteristics}}, in: \bibinfo{editor}{Konings, R.J.M.} (Ed.),
  \bibinfo{booktitle}{Comprehensive Nuclear Materials}.
  \bibinfo{publisher}{Elsevier}. volume~\bibinfo{volume}{2}, pp.
  \bibinfo{pages}{217--232}.
\newblock \DOIprefix\doi{10.1016/B978-0-08-056033-5.00015-X}.
\bibitem[{Ma and Dudarev(2019)}]{Ma2019}
\bibinfo{author}{Ma, P.W.}, \bibinfo{author}{Dudarev, S.L.},
  \bibinfo{year}{2019}.
\newblock \bibinfo{title}{{Symmetry-broken self-interstitial defects in
  chromium, molybdenum, and tungsten}}.
\newblock \bibinfo{journal}{Physical Review Materials} \bibinfo{volume}{3},
  \bibinfo{pages}{043606}.
\newblock \DOIprefix\doi{10.1103/PhysRevMaterials.3.043606}.
\bibitem[{Mandadapu et~al.(2014)Mandadapu, Jones and Zimmerman}]{Mandadapu2014}
\bibinfo{author}{Mandadapu, K.K.}, \bibinfo{author}{Jones, R.E.},
  \bibinfo{author}{Zimmerman, J.A.}, \bibinfo{year}{2014}.
\newblock \bibinfo{title}{On the microscopic definitions of the dislocation
  density tensor}.
\newblock \bibinfo{journal}{Mathematics and Mechanics of Solids}
  \bibinfo{volume}{19}, \bibinfo{pages}{744–757}.
\newblock \DOIprefix\doi{10.1177/1081286513486792}.
\bibitem[{Mason et~al.(2020)Mason, Das, Derlet, Dudarev, London, Yu, Phillips,
  Yang, Mizohata, Xu and Hofmann}]{Mason2020}
\bibinfo{author}{Mason, D.R.}, \bibinfo{author}{Das, S.},
  \bibinfo{author}{Derlet, P.M.}, \bibinfo{author}{Dudarev, S.L.},
  \bibinfo{author}{London, A.J.}, \bibinfo{author}{Yu, H.},
  \bibinfo{author}{Phillips, N.W.}, \bibinfo{author}{Yang, D.},
  \bibinfo{author}{Mizohata, K.}, \bibinfo{author}{Xu, R.},
  \bibinfo{author}{Hofmann, F.}, \bibinfo{year}{2020}.
\newblock \bibinfo{title}{{Observation of Transient and Asymptotic Driven
  Structural States of Tungsten Exposed to Radiation}}.
\newblock \bibinfo{journal}{Physical Review Letters} \bibinfo{volume}{125},
  \bibinfo{pages}{225503}.
\newblock \DOIprefix\doi{10.1103/PhysRevLett.125.225503}.
\bibitem[{Mason et~al.(2021)Mason, Granberg, Boleininger, Schwarz-Selinger,
  Nordlund and Dudarev}]{Mason2021}
\bibinfo{author}{Mason, D.R.}, \bibinfo{author}{Granberg, F.},
  \bibinfo{author}{Boleininger, M.}, \bibinfo{author}{Schwarz-Selinger, T.},
  \bibinfo{author}{Nordlund, K.}, \bibinfo{author}{Dudarev, S.L.},
  \bibinfo{year}{2021}.
\newblock \bibinfo{title}{Parameter-free quantitative simulation of high-dose
  microstructure and hydrogen retention in ion-irradiated tungsten}.
\newblock \bibinfo{journal}{Physical Review Materials} \bibinfo{volume}{5},
  \bibinfo{pages}{095403}.
\newblock \DOIprefix\doi{10.1103/PhysRevMaterials.5.095403}.
\bibitem[{Mendelev and Ackland(2007)}]{Mendelev2007}
\bibinfo{author}{Mendelev, M.I.}, \bibinfo{author}{Ackland, G.J.},
  \bibinfo{year}{2007}.
\newblock \bibinfo{title}{{Development of an interatomic potential for the
  simulation of phase transformations in zirconium}}.
\newblock \bibinfo{journal}{Philosophical Magazine Letters}
  \bibinfo{volume}{87}, \bibinfo{pages}{349--359}.
\newblock \DOIprefix\doi{10.1080/09500830701191393}.
\bibitem[{Milojevi{\'{c}}(2010)}]{Milojevic2010}
\bibinfo{author}{Milojevi{\'{c}}, S.}, \bibinfo{year}{2010}.
\newblock \bibinfo{title}{{Power law distributions in information science:
  Making the case for logarithmic binning}}.
\newblock \bibinfo{journal}{Journal of the American Society for Information
  Science and Technology} \bibinfo{volume}{61}, \bibinfo{pages}{2417--2425}.
\newblock \DOIprefix\doi{10.1002/asi.21426}.
\bibitem[{Nicodemus and Staub(1953)}]{Nicodemus1953}
\bibinfo{author}{Nicodemus, D.B.}, \bibinfo{author}{Staub, H.H.},
  \bibinfo{year}{1953}.
\newblock \bibinfo{title}{{Fission Neutron Spectrum of U$^{235}$}}.
\newblock \bibinfo{journal}{Physical Review} \bibinfo{volume}{89},
  \bibinfo{pages}{1288}.
\newblock \DOIprefix\doi{10.1103/PhysRev.89.1288}.
\bibitem[{Nordlund et~al.(2018)Nordlund, Zinkle, Sand, Granberg, Averback,
  Stoller, Suzudo, Malerba, Banhart, Weber, Willaime, Dudarev and
  Simeone}]{Nordlund2018}
\bibinfo{author}{Nordlund, K.}, \bibinfo{author}{Zinkle, S.J.},
  \bibinfo{author}{Sand, A.E.}, \bibinfo{author}{Granberg, F.},
  \bibinfo{author}{Averback, R.S.}, \bibinfo{author}{Stoller, R.},
  \bibinfo{author}{Suzudo, T.}, \bibinfo{author}{Malerba, L.},
  \bibinfo{author}{Banhart, F.}, \bibinfo{author}{Weber, W.J.},
  \bibinfo{author}{Willaime, F.}, \bibinfo{author}{Dudarev, S.L.},
  \bibinfo{author}{Simeone, D.}, \bibinfo{year}{2018}.
\newblock \bibinfo{title}{Improving atomic displacement and replacement
  calculations with physically realistic damage models}.
\newblock \bibinfo{journal}{Nature Communications} \bibinfo{volume}{9},
  \bibinfo{pages}{1084}.
\newblock \DOIprefix\doi{10.1038/s41467-018-03415-5}.
\bibitem[{Nye(1953)}]{Nye1953}
\bibinfo{author}{Nye, J.}, \bibinfo{year}{1953}.
\newblock \bibinfo{title}{{Some geometrical relations in dislocated crystals}}.
\newblock \bibinfo{journal}{Acta Metallurgica} \bibinfo{volume}{1},
  \bibinfo{pages}{153--162}.
\newblock \DOIprefix\doi{10.1016/0001-6160(53)90054-6}.
\bibitem[{Onimus and Bechade(2012)}]{Onimus2012}
\bibinfo{author}{Onimus, F.}, \bibinfo{author}{Bechade, J.L.},
  \bibinfo{year}{2012}.
\newblock \bibinfo{title}{{4.01 - Radiation Effects in Zirconium Alloys}}, in:
  \bibinfo{editor}{Konings, R.J.} (Ed.), \bibinfo{booktitle}{Comprehensive
  Nuclear Materials}. \bibinfo{publisher}{Elsevier}.
  volume~\bibinfo{volume}{4}, pp. \bibinfo{pages}{1--31}.
\newblock \DOIprefix\doi{10.1016/B978-0-08-056033-5.00064-1}.
\bibitem[{Onimus et~al.(2022)Onimus, Gélébart and Brenner}]{Onimus2022}
\bibinfo{author}{Onimus, F.}, \bibinfo{author}{Gélébart, L.},
  \bibinfo{author}{Brenner, R.}, \bibinfo{year}{2022}.
\newblock \bibinfo{title}{{Polycrystalline simulations of in-reactor
  deformation of recrystallized Zircaloy-4 tubes: Fast Fourier Transform
  computations and mean-field self-consistent model}}.
\newblock \bibinfo{journal}{International Journal of Plasticity}
  \bibinfo{volume}{153}, \bibinfo{pages}{103272}.
\newblock \DOIprefix\doi{10.1016/j.ijplas.2022.103272}.
\bibitem[{Paxton(2014)}]{Paxton2014}
\bibinfo{author}{Paxton, A.T.}, \bibinfo{year}{2014}.
\newblock \bibinfo{title}{From quantum mechanics to physical metallurgy of
  steels}.
\newblock \bibinfo{journal}{Materials Science and Technology}
  \bibinfo{volume}{30}, \bibinfo{pages}{1063}.
\newblock \DOIprefix\doi{10.1179/1743284714Y.0000000521}.
\bibitem[{Plimpton(1995)}]{Plimpton1995}
\bibinfo{author}{Plimpton, S.}, \bibinfo{year}{1995}.
\newblock \bibinfo{title}{{Fast Parallel Algorithms for Short-Range Molecular
  Dynamics}}.
\newblock \bibinfo{journal}{Journal of Computational Physics}
  \bibinfo{volume}{117}, \bibinfo{pages}{1--19}.
\newblock \DOIprefix\doi{10.1006/jcph.1995.1039}.
\bibitem[{Pomerance(1951)}]{Pomerance1951}
\bibinfo{author}{Pomerance, H.}, \bibinfo{year}{1951}.
\newblock \bibinfo{title}{{Thermal Neutron Capture Cross Sections}}.
\newblock \bibinfo{journal}{Physical Review} \bibinfo{volume}{83},
  \bibinfo{pages}{641--645}.
\newblock \DOIprefix\doi{10.1103/PhysRev.83.641}.
\bibitem[{Rib{\'{a}}rik et~al.(2020)Rib{\'{a}}rik, J{\'{o}}ni and
  Ung{\'{a}}r}]{Ribarik2020}
\bibinfo{author}{Rib{\'{a}}rik, G.}, \bibinfo{author}{J{\'{o}}ni, B.},
  \bibinfo{author}{Ung{\'{a}}r, T.}, \bibinfo{year}{2020}.
\newblock \bibinfo{title}{{The Convolutional Multiple Whole Profile (CMWP)
  Fitting Method, a Global Optimization Procedure for Microstructure
  Determination}}.
\newblock \bibinfo{journal}{Crystals} \bibinfo{volume}{10},
  \bibinfo{pages}{623}.
\newblock \DOIprefix\doi{10.3390/cryst10070623}.
\bibitem[{Rickover et~al.(1975)Rickover, Geiger and Lustman}]{Rickover1975}
\bibinfo{author}{Rickover, H.G.}, \bibinfo{author}{Geiger, L.D.},
  \bibinfo{author}{Lustman, B.}, \bibinfo{year}{1975}.
\newblock \bibinfo{type}{Technical Report}. Technical Information Center.
  \bibinfo{address}{U.S. Department of Energy}.
\newblock \DOIprefix\doi{10.2172/4240391}.
\bibitem[{Sand et~al.(2013)Sand, Dudarev and Nordlund}]{Sand2013}
\bibinfo{author}{Sand, A.E.}, \bibinfo{author}{Dudarev, S.L.},
  \bibinfo{author}{Nordlund, K.}, \bibinfo{year}{2013}.
\newblock \bibinfo{title}{High-energy collision cascades in tungsten:
  Dislocation loops structure and clustering scaling laws}.
\newblock \bibinfo{journal}{EPL} \bibinfo{volume}{103}, \bibinfo{pages}{46003}.
\newblock \DOIprefix\doi{10.1209/0295-5075/103/46003}.
\bibitem[{Schoeck(1962)}]{Schoeck1962}
\bibinfo{author}{Schoeck, G.}, \bibinfo{year}{1962}.
\newblock \bibinfo{title}{{Correlation between Dislocation Length and
  Density}}.
\newblock \bibinfo{journal}{Journal of Applied Physics} \bibinfo{volume}{33},
  \bibinfo{pages}{1745--1747}.
\newblock \DOIprefix\doi{10.1063/1.1728821}.
\bibitem[{Sears(2006)}]{Sears2006}
\bibinfo{author}{Sears, V.F.}, \bibinfo{year}{2006}.
\newblock \bibinfo{title}{{Neutron scattering lengths and cross sections}}.
\newblock \bibinfo{journal}{Neutron News} \bibinfo{volume}{3},
  \bibinfo{pages}{26--37}.
\newblock \DOIprefix\doi{10.1080/10448639208218770}.
\bibitem[{Seymour et~al.(2017)Seymour, Frankel, Balogh, Ung{\'{a}}r, Thompson,
  J{\"{a}}dern{\"{a}}s, Romero, Hallstadius, Daymond, Rib{\'{a}}rik and
  Preuss}]{Seymour2017}
\bibinfo{author}{Seymour, T.}, \bibinfo{author}{Frankel, P.},
  \bibinfo{author}{Balogh, L.}, \bibinfo{author}{Ung{\'{a}}r, T.},
  \bibinfo{author}{Thompson, S.}, \bibinfo{author}{J{\"{a}}dern{\"{a}}s, D.},
  \bibinfo{author}{Romero, J.}, \bibinfo{author}{Hallstadius, L.},
  \bibinfo{author}{Daymond, M.}, \bibinfo{author}{Rib{\'{a}}rik, G.},
  \bibinfo{author}{Preuss, M.}, \bibinfo{year}{2017}.
\newblock \bibinfo{title}{{Evolution of dislocation structure in neutron
  irradiated Zircaloy-2 studied by synchrotron {X}-ray diffraction peak profile
  analysis}}.
\newblock \bibinfo{journal}{Acta Materialia} \bibinfo{volume}{126},
  \bibinfo{pages}{102--113}.
\newblock \DOIprefix\doi{10.1016/j.actamat.2016.12.031}.
\bibitem[{Simmons and Baluffi(1958)}]{Simmons1958}
\bibinfo{author}{Simmons, R.O.}, \bibinfo{author}{Baluffi, R.W.},
  \bibinfo{year}{1958}.
\newblock \bibinfo{title}{{X-Ray Study of Deuteron-Irradiated Copper near
  10$^{\circ}$K}}.
\newblock \bibinfo{journal}{Physical Review} \bibinfo{volume}{109},
  \bibinfo{pages}{1142}.
\newblock \DOIprefix\doi{10.1103/PhysRev.109.1142}.
\bibitem[{Stoller et~al.(2013)Stoller, Toloczko, Was, Certain, Dwaraknath and
  Garner}]{Stoller2013}
\bibinfo{author}{Stoller, R.}, \bibinfo{author}{Toloczko, M.},
  \bibinfo{author}{Was, G.}, \bibinfo{author}{Certain, A.},
  \bibinfo{author}{Dwaraknath, S.}, \bibinfo{author}{Garner, F.},
  \bibinfo{year}{2013}.
\newblock \bibinfo{title}{On the use of {SRIM} for computing radiation damage
  exposure}.
\newblock \bibinfo{journal}{Nuclear Instruments and Methods in Physics Research
  Section B: Beam Interactions with Materials and Atoms} \bibinfo{volume}{310},
  \bibinfo{pages}{75--80}.
\newblock \DOIprefix\doi{10.1016/j.nimb.2013.05.008}.
\bibitem[{Stukowski(2010)}]{Stukowski2010b}
\bibinfo{author}{Stukowski, A.}, \bibinfo{year}{2010}.
\newblock \bibinfo{title}{{Visualization and analysis of atomistic simulation
  data with OVITO-the Open Visualization Tool}}.
\newblock \bibinfo{journal}{Modelling and Simulation in Materials Science and
  Engineering} \bibinfo{volume}{18}, \bibinfo{pages}{015012}.
\newblock \DOIprefix\doi{10.1088/0965-0393/18/1/015012}.
\bibitem[{Stukowski et~al.(2012)Stukowski, Bulatov and
  Arsenlis}]{Stukowski2012}
\bibinfo{author}{Stukowski, A.}, \bibinfo{author}{Bulatov, V.V.},
  \bibinfo{author}{Arsenlis, A.}, \bibinfo{year}{2012}.
\newblock \bibinfo{title}{{Automated identification and indexing of
  dislocations in crystal interfaces}}.
\newblock \bibinfo{journal}{Modelling and Simulation in Materials Science and
  Engineering} \bibinfo{volume}{20}, \bibinfo{pages}{085007}.
\newblock \DOIprefix\doi{10.1088/0965-0393/20/8/085007}.
\bibitem[{Terentyev et~al.(2014)Terentyev, Heinola, Bakaev and
  Zhurkin}]{Terentyev2014}
\bibinfo{author}{Terentyev, D.}, \bibinfo{author}{Heinola, K.},
  \bibinfo{author}{Bakaev, A.}, \bibinfo{author}{Zhurkin, E.E.},
  \bibinfo{year}{2014}.
\newblock \bibinfo{title}{Carbon–vacancy interaction controls lattice damage
  recovery in iron}.
\newblock \bibinfo{journal}{Scripta Materialia} \bibinfo{volume}{86},
  \bibinfo{pages}{9 -- 12}.
\newblock \DOIprefix\doi{10.1016/j.scriptamat.2014.04.003}.
\bibitem[{Theodorou et~al.(2022)Theodorou, Syskaki, Kotsina, Axiotis,
  Apostolopoulos and Fu}]{Apostolopoulos2022}
\bibinfo{author}{Theodorou, A.}, \bibinfo{author}{Syskaki, M.A.},
  \bibinfo{author}{Kotsina, Z.}, \bibinfo{author}{Axiotis, M.},
  \bibinfo{author}{Apostolopoulos, G.}, \bibinfo{author}{Fu, C.C.},
  \bibinfo{year}{2022}.
\newblock \bibinfo{title}{Interactions between irradiation defects and nitrogen
  in $\alpha$-{Fe}: an integrated experimental and theoretical study}.
\newblock \bibinfo{journal}{Acta Materialia} \bibinfo{volume}{239},
  \bibinfo{pages}{118227}.
\newblock \DOIprefix\doi{10.1016/j.actamat.2022.118227}.
\bibitem[{Tian et~al.(2021)Tian, Wang, Feng, Zheng, Liu and Zhou}]{Tian2021b}
\bibinfo{author}{Tian, J.}, \bibinfo{author}{Wang, H.}, \bibinfo{author}{Feng,
  Q.}, \bibinfo{author}{Zheng, J.}, \bibinfo{author}{Liu, X.},
  \bibinfo{author}{Zhou, W.}, \bibinfo{year}{2021}.
\newblock \bibinfo{title}{{Heavy radiation damage in alpha zirconium at
  cryogenic temperature: A computational study}}.
\newblock \bibinfo{journal}{Journal of Nuclear Materials}
  \bibinfo{volume}{555}, \bibinfo{pages}{153159}.
\newblock \DOIprefix\doi{10.1016/j.jnucmat.2021.153159}.
\bibitem[{Topping et~al.(2018)Topping, Ung{\'{a}}r, Race, Harte, Garner,
  Baxter, Dumbill, Frankel and Preuss}]{Topping2018}
\bibinfo{author}{Topping, M.}, \bibinfo{author}{Ung{\'{a}}r, T.},
  \bibinfo{author}{Race, C.P.}, \bibinfo{author}{Harte, A.},
  \bibinfo{author}{Garner, A.}, \bibinfo{author}{Baxter, F.},
  \bibinfo{author}{Dumbill, S.}, \bibinfo{author}{Frankel, P.},
  \bibinfo{author}{Preuss, M.}, \bibinfo{year}{2018}.
\newblock \bibinfo{title}{{Investigating the thermal stability of
  irradiation-induced damage in a zirconium alloy with novel in situ
  techniques}}.
\newblock \bibinfo{journal}{Acta Materialia} \bibinfo{volume}{145},
  \bibinfo{pages}{255--263}.
\newblock \DOIprefix\doi{10.1016/j.actamat.2017.11.051}.
\bibitem[{Ung{\'{a}}r et~al.(2021)Ung{\'{a}}r, Frankel, Rib{\'{a}}rik, Race and
  Preuss}]{Ungar2021b}
\bibinfo{author}{Ung{\'{a}}r, T.}, \bibinfo{author}{Frankel, P.},
  \bibinfo{author}{Rib{\'{a}}rik, G.}, \bibinfo{author}{Race, C.P.},
  \bibinfo{author}{Preuss, M.}, \bibinfo{year}{2021}.
\newblock \bibinfo{title}{{Size-distribution of irradiation-induced
  dislocation-loops in materials used in the nuclear industry}}.
\newblock \bibinfo{journal}{Journal of Nuclear Materials}
  \bibinfo{volume}{550}, \bibinfo{pages}{152945}.
\newblock \DOIprefix\doi{10.1016/j.jnucmat.2021.152945}.
\bibitem[{Ung\'{a}r et~al.(2021)Ung\'{a}r, Ribarik, Topping, Jones, Xu, Hulse,
  Harte, Tichy, Race, Frankel and Preuss}]{Ungar2021a}
\bibinfo{author}{Ung\'{a}r, T.}, \bibinfo{author}{Ribarik, G.},
  \bibinfo{author}{Topping, M.}, \bibinfo{author}{Jones, R.M.A.},
  \bibinfo{author}{Xu, X.D.}, \bibinfo{author}{Hulse, R.},
  \bibinfo{author}{Harte, A.}, \bibinfo{author}{Tichy, G.},
  \bibinfo{author}{Race, C.P.}, \bibinfo{author}{Frankel, P.},
  \bibinfo{author}{Preuss, M.}, \bibinfo{year}{2021}.
\newblock \bibinfo{title}{{Characterizing dislocation loops in irradiated
  polycrystalline Zr alloys by X-ray line profile analysis of powder
  diffraction patterns with satellites}}.
\newblock \bibinfo{journal}{Journal of Applied Crystallography}
  \bibinfo{volume}{54}, \bibinfo{pages}{803--821}.
\newblock \DOIprefix\doi{10.1107/S1600576721002673}.
\bibitem[{Ungár et~al.(1999)Ungár, Dragomir, Révész and
  Borbély}]{Ungar1999}
\bibinfo{author}{Ungár, T.}, \bibinfo{author}{Dragomir, I.},
  \bibinfo{author}{Révész, {\'A}.}, \bibinfo{author}{Borbély, A.},
  \bibinfo{year}{1999}.
\newblock \bibinfo{title}{The contrast factors of dislocations in cubic
  crystals: the dislocation model of strain anisotropy in practice}.
\newblock \bibinfo{journal}{Journal of Applied Crystallography}
  \bibinfo{volume}{32}, \bibinfo{pages}{992--1002}.
\newblock \DOIprefix\doi{10.1107/S0021889899009334}.
\bibitem[{Varvenne and Clouet(2017)}]{Varvenne2017}
\bibinfo{author}{Varvenne, C.}, \bibinfo{author}{Clouet, E.},
  \bibinfo{year}{2017}.
\newblock \bibinfo{title}{{Elastic dipoles of point defects from atomistic
  simulations}}.
\newblock \bibinfo{journal}{Physical Review B} \bibinfo{volume}{96},
  \bibinfo{pages}{224103}.
\newblock \DOIprefix\doi{10.1103/PhysRevB.96.224103}.
\bibitem[{Varvenne et~al.(2014)Varvenne, Mackain and Clouet}]{Varvenne2014}
\bibinfo{author}{Varvenne, C.}, \bibinfo{author}{Mackain, O.},
  \bibinfo{author}{Clouet, E.}, \bibinfo{year}{2014}.
\newblock \bibinfo{title}{{Vacancy clustering in zirconium: An atomic-scale
  study}}.
\newblock \bibinfo{journal}{Acta Materialia} \bibinfo{volume}{78},
  \bibinfo{pages}{65--77}.
\newblock \DOIprefix\doi{10.1016/j.actamat.2014.06.012}.
\bibitem[{Vineyard(1957)}]{Vineyard1957}
\bibinfo{author}{Vineyard, G.H.}, \bibinfo{year}{1957}.
\newblock \bibinfo{title}{Frequency factors and isotope effects in solid state
  rate processes}.
\newblock \bibinfo{journal}{J. Phys. Chem. Solids} \bibinfo{volume}{3},
  \bibinfo{pages}{121 -- 127}.
\newblock \DOIprefix\doi{10.1016/0022-3697(57)90059-8}.
\bibitem[{Wang et~al.(2023)Wang, Guo, Schwarz-Selinger, Yuan, Ge, Cheng, Zhang,
  Cao, Fu and Lu}]{Wang2022}
\bibinfo{author}{Wang, S.}, \bibinfo{author}{Guo, W.},
  \bibinfo{author}{Schwarz-Selinger, T.}, \bibinfo{author}{Yuan, Y.},
  \bibinfo{author}{Ge, L.}, \bibinfo{author}{Cheng, L.},
  \bibinfo{author}{Zhang, X.}, \bibinfo{author}{Cao, X.}, \bibinfo{author}{Fu,
  E.}, \bibinfo{author}{Lu, G.H.}, \bibinfo{year}{2023}.
\newblock \bibinfo{title}{Dynamic equilibrium of displacement damage defects in
  heavy-ion irradiated tungsten}.
\newblock \bibinfo{journal}{Acta Materialia} ,
  \bibinfo{pages}{118578}\DOIprefix\doi{10.1016/j.actamat.2022.118578}.
\bibitem[{Warwick et~al.(2021)Warwick, Boleininger and Dudarev}]{Warwick2021}
\bibinfo{author}{Warwick, A.R.}, \bibinfo{author}{Boleininger, M.},
  \bibinfo{author}{Dudarev, S.L.}, \bibinfo{year}{2021}.
\newblock \bibinfo{title}{{Microstructural complexity and dimensional changes
  in heavily irradiated zirconium}}.
\newblock \bibinfo{journal}{Physical Review Materials} \bibinfo{volume}{5},
  \bibinfo{pages}{113604}.
\newblock \DOIprefix\doi{10.1103/PhysRevMaterials.5.113604}.
\bibitem[{Wilkens(1970)}]{Wilkens1970}
\bibinfo{author}{Wilkens, M.}, \bibinfo{year}{1970}.
\newblock \bibinfo{title}{{Theoretical Aspects of Kinematical X-ray Diffraction
  Profiles from Crystals Containing Dislocation Distribution}}, in:
  \bibinfo{editor}{Simmons, J.A.}, \bibinfo{editor}{de~Wit, R.},
  \bibinfo{editor}{Bullough, R.} (Eds.), \bibinfo{booktitle}{Fundamental
  Aspects of Dislocation Theory}, \bibinfo{address}{Washington, D.C.}. pp.
  \bibinfo{pages}{1195--1221}.
\bibitem[{Willaime and Massobrio(1989)}]{Willaime1989}
\bibinfo{author}{Willaime, F.}, \bibinfo{author}{Massobrio, C.},
  \bibinfo{year}{1989}.
\newblock \bibinfo{title}{{Temperature-Induced hcp-bcc Phase Transformation in
  Zirconium: A Lattice and Molecular-Dynamics Study Based on an N-Body
  Potential}}.
\newblock \bibinfo{journal}{Physical Review Letters} \bibinfo{volume}{63},
  \bibinfo{pages}{2244--2247}.
\newblock \DOIprefix\doi{10.1103/PhysRevLett.63.2244}.
\bibitem[{Yi et~al.(2015)Yi, Sand, Mason, Kirk, Roberts, Nordlund and
  Dudarev}]{Yi2015}
\bibinfo{author}{Yi, X.}, \bibinfo{author}{Sand, A.E.}, \bibinfo{author}{Mason,
  D.R.}, \bibinfo{author}{Kirk, M.A.}, \bibinfo{author}{Roberts, S.G.},
  \bibinfo{author}{Nordlund, K.}, \bibinfo{author}{Dudarev, S.L.},
  \bibinfo{year}{2015}.
\newblock \bibinfo{title}{Direct observation of size scaling and elastic
  interaction between nano-scale defects in collision cascades}.
\newblock \bibinfo{journal}{EPL} \bibinfo{volume}{110}, \bibinfo{pages}{36001}.
\newblock \DOIprefix\doi{10.1209/0295-5075/110/36001}.
\bibitem[{Yu et~al.(2017)Yu, Yao, Idrees, Zhang, Kirk and Daymond}]{Yu2017}
\bibinfo{author}{Yu, H.}, \bibinfo{author}{Yao, Z.}, \bibinfo{author}{Idrees,
  Y.}, \bibinfo{author}{Zhang, H.K.}, \bibinfo{author}{Kirk, M.A.},
  \bibinfo{author}{Daymond, M.R.}, \bibinfo{year}{2017}.
\newblock \bibinfo{title}{Accumulation of dislocation loops in the $\alpha$
  phase of {Zr} {Excel} alloy under heavy ion irradiation}.
\newblock \bibinfo{journal}{Journal of Nuclear Materials}
  \bibinfo{volume}{491}, \bibinfo{pages}{232--241}.
\newblock \DOIprefix\doi{10.1016/j.jnucmat.2017.04.038}.
\bibitem[{Zarestky(1979)}]{Zarestky1979}
\bibinfo{author}{Zarestky, J.L.}, \bibinfo{year}{1979}.
\newblock \bibinfo{title}{Lattice dynamics of hcp and bcc zirconium}.
\newblock Ph.D. thesis. Iowa State University.
\newblock \DOIprefix\doi{10.31274/rtd-180813-3520}.
\bibitem[{Zhou et~al.(2006)Zhou, Jenkins, Dudarev, Sutton and Kirk}]{Zhou2006}
\bibinfo{author}{Zhou, Z.}, \bibinfo{author}{Jenkins, M.L.},
  \bibinfo{author}{Dudarev, S.L.}, \bibinfo{author}{Sutton, A.P.},
  \bibinfo{author}{Kirk, M.A.}, \bibinfo{year}{2006}.
\newblock \bibinfo{title}{Simulations of weak-beam diffraction contrast images
  of dislocation loops by the many-beam {Howie–Basinski} equations}.
\newblock \bibinfo{journal}{Philosophical Magazine} \bibinfo{volume}{86},
  \bibinfo{pages}{4851–4881}.
\newblock \DOIprefix\doi{10.1080/14786430600615041}.
\bibitem[{Ziegler et~al.(2010)Ziegler, Ziegler and Biersack}]{Ziegler2010}
\bibinfo{author}{Ziegler, J.F.}, \bibinfo{author}{Ziegler, M.D.},
  \bibinfo{author}{Biersack, J.P.}, \bibinfo{year}{2010}.
\newblock \bibinfo{title}{{SRIM - The stopping and range of ions in matter
  (2010)}}.
\newblock \bibinfo{journal}{Nuclear Instruments and Methods in Physics Research
  Section B: Beam Interactions with Materials and Atoms} \bibinfo{volume}{268},
  \bibinfo{pages}{1818--1823}.
\newblock \DOIprefix\doi{10.1016/J.NIMB.2010.02.091}.
\bibitem[{Zinkle and Was(2013)}]{Zinkle2013}
\bibinfo{author}{Zinkle, S.}, \bibinfo{author}{Was, G.}, \bibinfo{year}{2013}.
\newblock \bibinfo{title}{{Materials challenges in nuclear energy}}.
\newblock \bibinfo{journal}{Acta Materialia} \bibinfo{volume}{61},
  \bibinfo{pages}{735--758}.
\newblock \DOIprefix\doi{10.1016/j.actamat.2012.11.004}.

\end{thebibliography}



\end{document}